\documentclass[journal,10pt]{IEEEtran}
\normalsize

\usepackage{cite}
\usepackage{graphicx}

\usepackage[cmex10]{amsmath}
\usepackage{amssymb}
\usepackage{booktabs}
\usepackage{threeparttable}
\usepackage{algorithm}
\usepackage{algpseudocode}
\usepackage{amsfonts}
\usepackage{xcolor}
\usepackage{color}

\usepackage{tabularx} %

\usepackage{caption}

\usepackage{arcs}

\usepackage{enumitem}
\usepackage{framed}

\usepackage{dblfloatfix} %

\usepackage{soul}

\usepackage{cancel} %

\usepackage{verbatim}%

\usepackage{makecell}

\usepackage{empheq} %

\usepackage{underoverlap} %

\usepackage{hyperref}
\hypersetup{pdftex,colorlinks=true,allcolors=black}

\usepackage{calc}

\usepackage{textcomp}
\def\BibTeX{{\rm B\kern-.05em{\sc i\kern-.025em b}\kern-.08em
		T\kern-.1667em\lower.7ex\hbox{E}\kern-.125emX}}

\newtheorem{theorem}{\bf  Theorem}

\newtheorem{goal}{\bf Goal}

\newtheorem{corollary}{\bf  Corollary}

\newtheorem{lemma}{\bf  Lemma}

\newcommand{\mA}{\mathsf{A}}
\newcommand{\mB}{\mathsf{B}}
\newcommand{\mC}{\mathsf{C}}
\newcommand{\mD}{\mathsf{D}}
\newcommand{\mE}{\mathsf{E}}
\newcommand{\mF}{\mathsf{F}}
\newcommand{\mG}{\mathsf{G}}
\newcommand{\mH}{\mathsf{H}}
\newcommand{\mI}{\mathsf{I}}
\newcommand{\mJ}{\mathsf{J}}
\newcommand{\mK}{\mathsf{K}}
\newcommand{\mL}{\mathsf{L}}

\newcommand{\mN}{\mathsf{N}}
\newcommand{\mO}{\mathsf{O}}
\newcommand{\mP}{\mathsf{P}}
\newcommand{\mS}{\mathsf{S}}

\newcommand{\mQ}{\mathsf{Q}}

\newcommand{\mR}{\mathsf{R}}

\newcommand{\mj}{\mathsf{j}}

\newcommand{\epa}{\epsilon_{\mathrm{a}}}
\newcommand{\epp}{\epsilon_{\mathrm{p}}}
\newcommand{\epr}{\epsilon_{\mathrm{r}}}

\newcommand{\myBigO}[1]{\mathcal{O}\left\{#1\right\}}

\newcommand{\myDef}{\overset{\Delta}{=}}
\newcommand{\myEqualOverset}[1]{\overset{#1}{=}}

\newcommand{\myAbs}[1]{\left|#1\right|}

\newcommand{\myLambdaMax}[1]{\lambda_{\mathrm{max}}\left\{#1\right\}}
\newcommand{\myLambdaMin}[1]{\lambda_{\mathrm{min}}\left\{#1\right\}}
\newcommand{\myVec}[1]{\mathrm{vec}\left\{#1\right\}}

\newcommand{\myEigen}[1]{\mathrm{eig}\left\{#1\right\}}

\newcommand{\myBracketRnd}[1]{\left(#1\right)}
\newcommand{\myBracketBig}[1]{\left\{#1\right\}}
\newcommand{\myBracketSqr}[1]{\left[#1\right]}

\newcommand{\myVecArrow}[1]{\vec{\mathsf{#1}}}

\newcommand{\myDiag}[1]{\mathrm{diag}\left(#1\right)}

\newcommand{\pcc}{periodic cross-correlation}

\newcommand{\myRound}[1]{\left\lfloor #1 \right\rceil}

\newcommand{\myModulo}[2]{\left\langle#1\right\rangle_{#2}}

\newcommand{\mySpaceTwoMM}{\vspace{2mm}}

\begin{document}

\title{Joint Communications and Sensing Employing Optimized MIMO-OFDM Signals
}

\author{
	Kai Wu,
	 J. Andrew Zhang,~\IEEEmembership{Senior Member,~IEEE}, Zhitong Ni, %
	Xiaojing Huang,~\IEEEmembership{Senior Member,~IEEE} \\
	Y. Jay Guo,~\IEEEmembership{Fellow,~IEEE}, and Shanzhi Chen,~\IEEEmembership{Fellow,~IEEE}
	\thanks{K. Wu, J. A. Zhang, Z. Ni, X. Huang and Y. J. Guo are with the Global Big Data Technologies Centre, University of Technology Sydney, Sydney, NSW 2007, Australia (e-mail: kai.wu@uts.edu.au; andrew.zhang@uts.edu.au; zhitong.ni@uts.edu.au; xiaojing.huang@uts.edu.au; jay.guo@uts.edu.au).}
	
	\thanks{Shanzhi Chen is with the State Key Lab of
		Wireless Mobile Communication, China Academy of Telecommunication
		Technology, Beijing China (email: chensz@cict.com).}

}

\maketitle

\begin{abstract}	
	Joint communication and sensing (JCAS) has the potential to improve the overall energy, cost and frequency efficiency of IoT systems. As a first effort, we propose to optimize the MIMO-OFDM data symbols carried by sub-carriers for better time- and spatial-domain signal orthogonality. This not only boosts the availability of usable signals for JCAS, but also significantly facilitates Internet-of-Things (IoT) devices to perform high-quality sensing.
	We establish an optimization problem that modifies data symbols on sub-carriers to enhance the above-mentioned signal orthogonality. We also develop an efficient algorithm to solve the problem based on the majorization-minimization framework. Moreover, we discover unique signal structures and features from the newly modeled problem, which substantially reduce the complexity of majorizing the objective function. We also develop new projectors to enforce the feasibility of the obtained solution. Simulations show that, compared with the 
	original communication waveform to achieve the same sensing performance, the optimized waveform can reduce the signal-to-noise ratio (SNR) requirement by $ 3\sim 4.5 $ dB, while the SNR loss for the uncoded bit error rate is only $ 1\sim1.5 $ dB.

\end{abstract}


\section{Introduction}
Joint communications and sensing (JCAS), aka, integrated sensing and communications (ISAC), have attracted extensive attention in Internet-of-Things (IoT) community recently \cite{Fan_ISAC4IoT2021networkIEEE,IoT_JCAS_OFDMcodeDivision2021,Kai_rahman2020enablingSurvey,IoT_LFM_MPSK2021,IIoT_droneskumar2021internet,IoT_RadarCom6GMassiveIoT2021}. 
\textit{On one hand}, it is because microwave sensing, which may also be known as radar/radio sensing or wireless sensing, has found extensive applications in numerous IoT use cases including intelligent transport, smart cities/homes/farms/health care, and human-activity sensing etc. \cite{Fan_ISAC4IoT2021networkIEEE,Sensing_reviewApplications2021sensorsJournal,JCAS_convegence6G_Jan2021Access}. 
\textit{On the other hand}, performing communications and sensing with one set of hardware and software resources can greatly improve energy-, cost- and frequency-efficiency \cite{IoT_JCAS_OFDMcodeDivision2021}. These efficiencies are of critical significance to the sustainable development of IoT systems, as they increasingly penetrate almost all aspects of our lives.

There are mainly two types of JCAS: passive and active \cite{Kai_rahman2020enablingSurvey,ni2021uplink}. We consider here the active JCAS that integrates sensing into a communication system with tolerable changes made to the latter.
So far, such active JCAS has been performed through designing dual-functional waveforms in spatial, time and frequency domains. In the spatial-domain,  dual-functional precoders/beamformers are generally designed to, e.g., approach desired sensing waveforms subject to signal-to-interference-plus-noise ratio (SINR) requirements for multi-user downlink multiple-input and multiple-output (MIMO) communications \cite{8386661}. 
In time and frequency domains, existing works mainly resort to designing the frame structure \cite{Cui18mat}, sub-carrier occupation \cite{DFRC_OFDMweightOptimization2020Infocom}, power allocation \cite{yuan2020waveform} and pilot/preamble signals \cite{Kumari20}. Communication data waveforms are rarely optimized, yet have been widely used, for JCAS. 

In \cite{DFRC_dsss2011procIeee}, a classical sensing method using the orthogonal frequency-division multiplexing (OFDM) communication signals is developed. The method transforms the echo signal into the frequency domain, removes the data symbols through point-wise divisions, and generates the range-Doppler map through a two-dimensional Fourier transform. 
The method has been extensively used in the past decade, particularly in automotive sensing \cite{DFRC_automotive2020SPmag,OFDM_autonomousDriv2019microwaveMag}. In \cite{DFRC_SC_OFDM}, the single-carrier OFDM (SC-OFDM) sensing is developed using communication data signals. The sensing is performed in a similar way to that in \cite{DFRC_dsss2011procIeee}. However, since SC-OFDM can have severely fluctuating amplitudes in the frequency domain, the point-wise division in \cite{DFRC_dsss2011procIeee} is replaced with the point-wise product to reduce noise enhancement.
In \cite{Kai_wu2021integrating_JSAC}, the methods \cite{DFRC_dsss2011procIeee,DFRC_SC_OFDM} are improved by relieving the constraints imposed by communication systems on sensing. In \cite{Kai_otfs_IoTindustrial,OTFS_jcas2020twc,OTFS_keskin2021radarTimeDomainICIisi,OTFS_radarRaviteja2019conference,OTFS_yuan2021integratedSensingCOmmunicationOTFS}, sensing using the data signals of orthogonal time-frequency space (OTFS) communication systems is developed. 

The communication data signals-based JCAS designs \cite{DFRC_dsss2011procIeee,DFRC_SC_OFDM,Kai_wu2021integrating_JSAC,Kai_otfs_IoTindustrial,OTFS_jcas2020twc,OTFS_keskin2021radarTimeDomainICIisi,OTFS_radarRaviteja2019conference} reviewed above mainly use a single antenna. The method \cite{DFRC_dsss2011procIeee} is 
later extended for MIMO cases. 
To ensure orthogonality of the signals transmitted by different antennas, an equidistant sub-carrier interleaving scheme is developed in \cite{DFRC_interleaveMIMO_OFDM_sturm2013spectrally}. The scheme lets antenna $ m $ uses sub-carriers $ m+iM $ for $ m=0,1,\cdots,M-1 $, where $ M $ is antenna number, and $ i\ge 0 $ such that $ (m+iM)~(\forall m,i) $ is no greater than the overall sub-carrier number. Considering that the equidistant interleaving can reduce the unambiguously measurable distance, the work \cite{Hakobyan20} develops a non-equidistant sub-carrier interleaving scheme. However, these interleaving schemes \cite{DFRC_interleaveMIMO_OFDM_sturm2013spectrally,Hakobyan20} can reduce communication spectral efficiency, as many sub-carriers need to be kept unused on each antenna.

In this work, we perform JCAS by optimizing the data signals of a MIMO-OFDM communication system. Data signal is used due to its much greater availability than other communication signals, such as preambles.
{Similar to \cite{DFRC_interleaveMIMO_OFDM_sturm2013spectrally,Hakobyan20}, we aim to enhance the orthogonality of data signals from all antennas\footnote{{Note that the sub-carrier interleaving scheme developed in \cite{DFRC_interleaveMIMO_OFDM_sturm2013spectrally} also leads to the orthogonality of signals from all antennas in the time domain. This because the link of the two domains in OFDM is a discrete Fourier transform (DFT) matrix which is unitary and does not change signal orthogonality.}}. 
	This is also to create a sensing scheme analogous to conventional orthogonal MIMO radars \cite{book_li2008mimo}. 
	Such a sensing scheme can be quite attractive to IoT systems, as a single transmission can illuminate a large spatial region with relatively uniform power distribution. Moreover, the waveform orthogonality can be exploited to virtually extend the array aperture (and hence the spatial resolution) to be much larger than the physical aperture that is typically small in IoT devices.  
	
	Different from \cite{DFRC_interleaveMIMO_OFDM_sturm2013spectrally,Hakobyan20}, we do not perform the optimization through sub-carrier interleaving (which, as mentioned earlier, will reduce communication spectral efficiency). Instead, we consider a more common case in MIMO-OFDM communications, where all antennas occupy the same sub-carriers \cite{book_mimoOFDMmatlab,book_ahmadi2019_5G}. Moreover, \textit{we aim to introduce minimal changes to data symbols carried by all sub-carriers to enhance the orthogonality of the corresponding time-domain signals from all antennas.}
	Data symbols are in the frequency domain and need to be strictly constrained to 
	maintain satisfactory communication performance. Thus, the proposed waveform design is distinct from the conventional MIMO radar waveform design that is directly performed and constrained in the time domain; see e.g., \cite{MM_JXSong_weightedPeak2016TSP,MM_unified_LCZhao_2017TSP,book_cui2020radarWaveformOptimization,MM_sequenceSetDesignJXSong2016TSP}.} 
The main contributions of our work are summarized as follows. 

\begin{enumerate}
	\item We propose to optimize the frequency-domain data symbols of a MIMO-OFDM communication system to enhance the time and spatial orthogonality of the signals transmitted by all antennas. To the best our knowledge, 
	this is the first work exploring the data symbol domain for JCAS waveform optimization. 
	We model an optimization problem for the proposed waveform design. In particular, we stack all data symbols over sub-carriers antenna-by-antenna into a long signal sequence; unitedly express all the cyclic auto- and cross-correlations of multi-antenna time-domain signals as functions of the same long signal sequence; and minimize the peak sidelobe of the correlations subject to the power and similarity constraints.

	\item  We develop an efficient algorithm to solve the optimization problem using the majorization-minimization (MM) framework \cite{MM_SP_Com_ML2017TSP_Babu}.
As suggested by the name, MM includes two stages of first majorizing the objective function and then minimizing the majorized function subject to certain constraints. For the first stage, we discover the unique signal structures and features from the newly modeled optimization problem, substantially reducing the computational complexity in majorizing the objective function. Take a typical IoT system with four antenna and $ 128 $ sub-carriers for instance; we reduce the majorization complexity from $ \myBigO{10^8} $ to $ \myBigO{10^2} $, significantly facilitating IoT devices to perform JCAS.  More details will be given in Section \ref{subsec: problem formulation}.
For the second stage of MM, we develop new projectors for two of the most common communication constellations, i.e., phase shift keying (PSK) and quadrature amplitude modulation (QAM), effectively confining the optimized solutions in feasible regions.

\end{enumerate}

We perform extensive 
simulations to validate the proposed waveform optimization by illustrating its impact on both communications and sensing performances. Key metrics include the uncoded bit error rate (BER) for communications and the detecting probability for sensing. 
For 4/8/16PSK modulations, the proposed waveform optimization can reduce the signal-to-noise ratio (SNR) requirement by more than $ 3 $ dB, compared with the original communication waveform, to achieve the same detecting probability that is greater than $ 0.75 $. Correspondingly, the BER loss is less than $ 1.5 $ dB. For the 16QAM modulation, our design can reduce the SNR requirement by up to $ 4.5 $ dB than using the original communication waveform, to achieve the same detecting probability of $ 0.85 $, while the BER loss is only up to $ 1.2 $ dB.

\textit{Paper Structure:} The rest of the paper starts with illustrating the system model, motivation and problem formulation in Section \ref{sec: system model}. The solution to the problem is developed by first majorizing the objective function in Section \ref{sec: majorization}, then developing projectors in Section \ref{sec: constraint projection} to enforce the feasibility of the solution, and finally establishing the overall algorithm for the proposed waveform optimization in Section \ref{sec: over all MM algorithm} with complexity analysis. Simulation results are provided in Section \ref{sec: simulations}, and the paper is concluded in Section \ref{sec: conclusions}.

\textit{Notations and symbols:} Throughout the paper, we use bold upper-case letters, e.g., $ \mathbf{X} $, for matrices; bold lower-case, e.g., $ \mathbf{x} $, for vectors; calligraphic upper-case letters, e.g., $ \mathcal{C} $, for sets; $ \mC $ for optimization-independent constants and $ \mathbb{C} $ denotes the set of complex numbers. We use $ (\cdot)^{\mathrm{T}} $ for transpose, $ (\cdot)^* $ for conjugation, $ (\cdot)^{\mathrm{H}} $ for conjugate transpose, and $ (\cdot)^{\dagger} $ for pseudo-inverse. Moreover, $ \odot $ is for the point-wise product, and $ \otimes $ for the Kronecker product. We let $ \Re\{x\} $ ($ \Im\{x\} $) take the real (imaginary) part of a complex number $ x $. 

While $ \myBracketSqr{\cdot}_{m,n=0}^{N-1} $ generates a square matrix with the row index $ m $, the column index $ n $ and the size $ N $, 
$ \myBracketSqr{\mathbf{x}}_n $ takes the $ n $-th entry of $ \mathbf{x} $; $ \myBracketSqr{\mathbf{X}}_{a,b} $ the $ (a,b) $-th entry of $ \mathbf{X} $; $ \myBracketSqr{\mathbf{X}}_{a,:} $ the $ a $-th row; $ \myBracketSqr{\mathbf{X}}_{:,b} $ the $ b $-th column.
The operation $ \mathrm{vec}\{\mathbf{X}\} $ vectorizes $ \mathbf{X} $ column-by-column; $ \myLambdaMax{\mathbf{X}} $ returns the maximum eigenvalue 
of $ \mathbf{X} $; $ \mathrm{diag}\{\mathbf{x}\} $ generates a diagonal matrix with $ \mathbf{x} $ put on the diagonal; $ \mathrm{diag}\{\mathbf{X}\} $ takes the diagonal entries; $ \mathbf{I}_x $ for an $ x $-size identity matrix.

Common symbols include: $ N $, $ M $, $ N_{\mathrm{CP}} $, $ N_{\mathrm{t}} $ denote the numbers of OFDM sub-carriers communication-transmitting antennas, the CP length, 
and the target number, respectively.

\section{System Model and Problem Formulation}\label{sec: system model}

In this section, we start with illustrating the system model and the rationale of the proposed waveform design. Then we formulate the optimization problem for the intended design. 

\subsection{System Model}

\begin{figure}[!t]
	\centering
	\includegraphics[width=88mm]{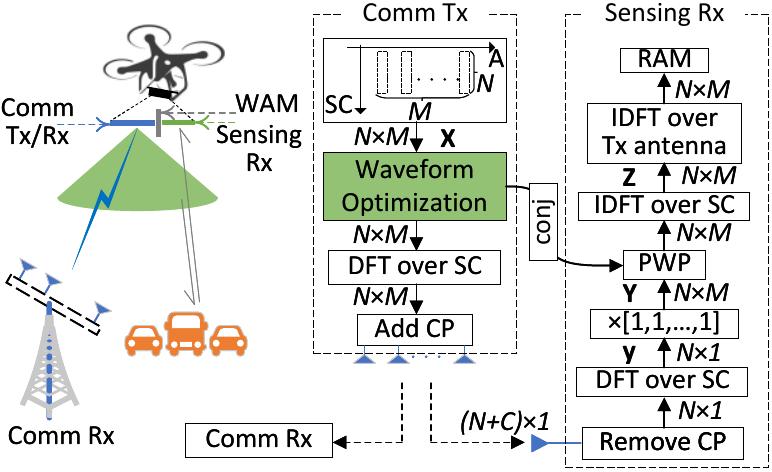}
	\caption{An exemplary scenario of the JCAS considered in this work, where the communication (comm) transmitter (Tx) performs point-to-point MIMO-OFDM communications with a communication receiver (Rx). The communication data signals are also used for sensing a large spatial region without any prior information on targets. 
		To improve the performance of such sensing, we propose to optimize the data symbols over $ N $ sub-carriers (SC) and $ M $ antennas (A), as will be developed in Sections \ref{sec: majorization}-\ref{sec: over all MM algorithm}. 
	The cyclic prefix (CP) is used to maintain sub-carrier orthogonality at the comm Rx. Key processing steps at a sensing Rx are illustrated, where PWP stands for point-wise product, RAM for range-angle map, and `conj' is short for `conjugate'. 
More details are given in Section \ref{subsec: sensing receiver processing and movitations}. 
	}
	\label{fig: JCAS model and tx rx processing}
\end{figure}

The considered JCAS is exemplified in Fig. \ref{fig: JCAS model and tx rx processing}. Point-to-point MIMO communications happen between a an unmanned aerial vehicle (UAV) and a ground-based access point. While this is only an example, the proposed method can be applied to similar communications in other IoT applications. The communication transmitter is equipped with multiple antennas transmitting independent data signals. 
The signals are also used for sensing. To do so, a sensing receiver, as co-located with the communication transmitter, is used to collect echo signals. The above configuration makes it reasonable to assume that the communication transmitter and sensing receiver are clock synchronized. 
In this work, we also assume that proper means, e.g., inserting wave-absorbing material (WAM) as illustrated in Fig.  \ref{fig: JCAS model and tx rx processing}, are employed to suppress the signal leakage form the transmitter to the receiver. Effectively addressing this problem may need more 
comprehensive transceiver designs \cite{FD_mmWave2021WirelessComm,FD_JCAS_enabler}, which are out of the scope of this work.

Let $ M $ denote the number of antennas at the communication transmitter, $ N $ the number of OFDM sub-carriers, and $ \mathbf{x}_m\in\mathbb{C}^{N\times 1} $ collect the $ N $ data symbols on the $ m $-th antenna. The data symbols are independently drawn from a constellation, such as QPSK. 
As shown in Fig. \ref{fig: JCAS model and tx rx processing}, the data symbols are stacked column-by-column into an $ N\times M $ matrix, as denoted by $ \mathbf{X}=\myBracketSqr{\mathbf{x}_0,\mathbf{x}_1,\cdots,\mathbf{x}_{M-1}} $. Consider the conventional communication transmission for now, and hence the waveform optimization shown in the figure is skipped. Then, a DFT
is taken over sub-carriers, i.e., each column of $ \mathbf{X} $. Denote the result by $ \mathbf{FX}\in \mathbb{C}^{N\times M} $, where $ \mathbf{F} $ 
is an $ N $-dimensional DFT matrix, as given by
\begin{align} \label{eq: F bf DFT matrix}
	\mathbf{F} = \myBracketSqr{e^{-\mj \frac{2\pi nm }{N}}}_{m,n=0}^{N-1}.
\end{align}
Here, $ \myBracketSqr{\cdot}_{m,n=0}^{N-1} $ generates an $ N\times N $ square matrix with the row index $ m $ and column index $ n $. Next, adding CP by copying the bottom $ N_{\mathrm{CP}} $ rows of $ \mathbf{FX} $ to its top. 
Each column of signal samples are then transmitted by an individual antenna after being processed by a radio frequency chain (DAC, frequency conversion and power amplification). The signals go through a communication (sensing) channel and arrive at a communication (sensing) receiver. 
Next, we briefly describe the signal processing steps at the two different receivers. 

\subsection{Communication Receiver Processing} \label{subsec: cmm receiver processing}
We describe the MIMO-OFDM communications at the receiver, for the purpose of introducing some communication performance metrics. These metrics will be observed later to evaluate the impact of the proposed design on data communications. To serve the above purpose in a concise yet sufficient manner, we consider a digital MIMO communication receiver that has $ K $ antennas. Since $ M $ independent data streams are transmitted, $ K\ge M $ is necessary for spatial multiplexing.

The signals received at all antennas are digitized by the receiver RF chains. Removing CPs and taking the DFT over $ N $ samples of each antenna, we can write the received signal matrix as $ \mathbf{H}\mathbf{X}^{\mathrm{T}} + \mathbf{N}(\in\mathbb{C}^{K\times N}) $, where $ \mathbf{H}(\in\mathbb{C}^{K\times M}) $ is the channel matrix and $ \mathbf{N} $ is the noise matrix. As is typical, $ [\mathbf{N}]_{k,n}\sim \mathcal{CN}\myBracketRnd{0,\sigma_{\mathrm{n}}^2 }~\forall k,n $, where $ [\cdot]_{k,n} $ takes the $ (k,n) $-th entry of a matrix, and $ \sigma_{\mathrm{n}}^2 $ denotes noise power. 
Applying the zero forcing combiner, the data symbol matrix can be estimated as $ \hat{\mathbf{X}}^{\mathrm{T}}=\mathbf{H}^{\dagger} \mathbf{H}\mathbf{X}^{\mathrm{T}} + \mathbf{H}^{\dagger}\mathbf{N} $.  
Demodulation can then be performed based on $ \hat{\mathbf{X}} $.

\subsection{Sensing Receiver Processing and Motivation} \label{subsec: sensing receiver processing and movitations}
For ease of exposition, we consider a single-antenna sensing receiver and illustrate the receiving steps based on a single OFDM symbol. 
After some typical RF processing, such as low-noise power amplification, frequency conversion and ADC, the CP in the digitized samples is first removed. Assuming that all target delays are smaller than the CP duration, an $ N $-dimensional DFT can be taken over the time-domain samples to obtain the following signals over sub-carriers,
\begin{align} \label{eq: y bf echo freq-domain}
	\mathbf{y} = \sum_{i=0}^{N_{\mathrm{t}}-1} \alpha_i \big( \mathbf{X} \mathbf{a}(\theta_i) \big) \odot \mathbf{b}(\tau_i) + \mathbf{n},
\end{align} 
where $ i $ denotes the target index, $ N_{\mathrm{t}} $ the total target number, $ \alpha_i $ absorbs the scattering coefficient and RF processing gain, $ \theta_i $ denotes the angle-of-departure (AoD) of the $ i $-th target, $ \mathbf{a}(\theta_i) $ the spatial steering vector, $ \tau_i $ the target delay, $ \mathbf{b}(\tau_i) $ the range steering vector, and $ \mathbf{n} $ a complex Gaussian noise vector with identically and independently distributed entries. The $ m $-th entry of $ \mathbf{a}(\theta_i) $ 
and the $ n $-th of $ \mathbf{b}(\tau_i) $ can be, respectively, given by 
\begin{align} \label{eq: a bf b bf steering vectors}
	a_{mi}\myDef \myBracketSqr{\mathbf{a}(\theta_i)}_m = e^{-\mj \pi m \sin\theta_i};~\myBracketSqr{\mathbf{b}(\tau_i)}_n=e^{\frac{-\mj 2\pi n \tau_i T_{\mathrm{s}}}{N} },
\end{align}
where the antenna spacing is half the wavelength, and $  T_{\mathrm{s}} $ denotes the sampling interval.

As shown in Fig. \ref{fig: JCAS model and tx rx processing}, stacking the $ M $ copies of $ \mathbf{y} $ in the row dimension yields a matrix $ \mathbf{Y} $.
Performing a point-wise product between $ \mathbf{Y} $ and $ \mathbf{X}^* $ and then taking the inverse DFT (IDFT) of each column, we obtain $ \mathbf{Z}=\mathrm{IDFT}(\mathbf{Y} \odot \mathbf{X}^*) $, where $ \mathbf{X} $ is the matrix of all data symbols over sub-carriers (rows) and antennas (columns) at the transmitter. The $ m $-th column of $ \mathbf{Z} $, as denoted by $ \mathbf{z}_m $, is the cyclic cross-correlation (CCC) between $ \mathrm{IDFT}(\mathbf{y}) $ and $ \mathrm{IDFT}(\mathbf{x}_m) $ \cite{DFRC_SC_OFDM}. 
Based on (\ref{eq: y bf echo freq-domain}), we can express $ \mathbf{z}_m $ as
\begin{align} \label{eq: z bf m}
	& \mathbf{z}_m = \sum_{i=0}^{N_{\mathrm{t}}-1} \alpha_i a_{mi} \mathrm{IDFT}\myBracketRnd{|\mathbf{x}_m|^2 \odot \mathbf{b}(\tau_i)} \nonumber\\
	&  + \sum_{\substack{m'=0\\m'\ne m}}^{M-1}\sum_{i=0}^{N_{\mathrm{t}}-1} \alpha_i a_{m'i} \mathrm{IDFT}\myBracketRnd{ \mathbf{x}_{m'} \odot \mathbf{x}_{m}^* \odot \mathbf{b}(\tau_i) } + \mathbf{n} ,
\end{align}
where $ a_{xi}~(x=m,m') $ is the $ x $-th entry of $ \mathbf{a}(\theta_i) $.

Based on (\ref{eq: a bf b bf steering vectors}) and the circular shift property of IDFT \cite{book_oppenheim1999discrete}, the shape of $ \mathrm{IDFT}\myBracketRnd{|\mathbf{x}_m|^2 \odot \mathbf{b}(\tau_i)} $ is only dependent on $ |\mathbf{x}_m|^2 $, while the peak location changes with $ \tau_i $. If two targets locate closely (i.e., their $ \tau_i $'s are similar), they interfere with each other. To reduce the interference, we expect $ \mathrm{IDFT}\myBracketRnd{|\mathbf{x}_m|^2 }~(\forall m) $ has low sidelobes. This leads to the first goal of waveform design, i.e., 

\begin{goal} \label{gl: low side lobes auto}
	{\it To reduce inter-target interference, $ \mathrm{IDFT}(\mathbf{x}_m) $ needs to have low sidelobes in its cyclic auto-correlation.} 
\end{goal}

The second term on the right-hand side of (\ref{eq: z bf m}) is the interference to antenna $ m $ caused by signals transmitted by other antennas. 
To reduce this interference, we expect $ \mathrm{IDFT}\myBracketRnd{ \mathbf{x}_{m'} \odot \mathbf{x}_{m}^*  } $ is small. This results in another waveform design goal:

\begin{goal} \label{gl: small ccc}
	{\it To reduce inter-antenna interference, $ \mathrm{IDFT}(\mathbf{x}_m)$ $(\forall m) $ and $ \mathrm{IDFT}(\mathbf{x}_{m'})~(\forall m'\ne m) $ need to have low CCC.} 
\end{goal}

Since $ \mathrm{IDFT}\myBracketRnd{|\mathbf{x}_m|^2 \odot \mathbf{b}(\tau_i)} $ accumulates at a target location, e.g., $ \myRound{\tau_i T_{\mathrm{s}}} $, in an approximately coherent manner, the $ \myRound{\tau_i T_{\mathrm{s}}} $-th entry of
$ \mathbf{z}_m $ differs over $ m $ mainly because of $ a_{mi} $, as seen from (\ref{eq: a bf b bf steering vectors}). Thus, taking another IDFT over each row of $ \mathbf{Z} $, the row dimension will then be transformed into the angular domain. Consequently, we obtain a range-angle map (RAM) that can be used for target detection and localization. To highlight the importance of the two waveform design goals, let us compare the 
RAMs obtained by different waveforms.

\begin{figure}[!t]
	\centering
	\includegraphics[width=88mm]{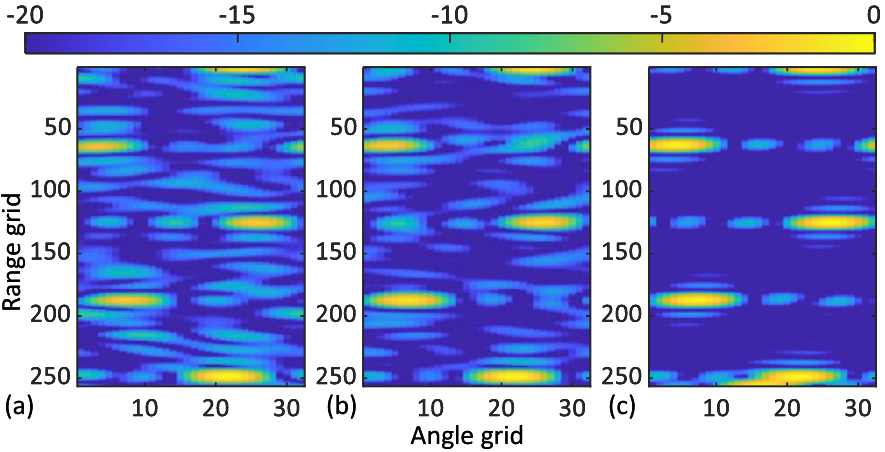}
	\caption{RAMs obtained by the original communication waveform using QPSK (a), the optimized one (using the methods to be developed) (b), and the orthogonal waveform designed in \cite{DFRC_interleaveMIMO_OFDM_sturm2013spectrally}. 
	Here, $ N=128 $, $ M=4 $ and $ N_{\mathrm{CP}}=32 $. Moreover, five targets are simulated with the range bins, i.e., $ \myRound{\tau_i T_{\mathrm{s}}} $ in (\ref{eq: a bf b bf steering vectors}), set as $ 1,8.75,16.5, 24.25$ and $   32 $. Their AoDs are randomly generated in $ [0,2\pi] $ radians. 
	The sizes of IDFTs over SC and antenna for generating RAM; see Fig. \ref{fig: JCAS model and tx rx processing}, are $ 8 $ times their actucal sizes. To highlight the impact of waveforms on sensing, noise is not added.}
	\label{fig: range angle map QPSK}
\end{figure}

Fig. \ref{fig: range angle map QPSK} plots the RAMs obtained using the original communication waveform in (a), the waveform optimized based on Goals \ref{gl: low side lobes auto} and \ref{gl: small ccc} in (b), and the orthogonal waveform in (c), where the optimization methods will be developed shortly.
The orthogonal waveform alternates the allocations of sub-carriers over antennas such that each sub-carrier is only used by a single antenna in an OFDM symbol \cite{DFRC_interleaveMIMO_OFDM_sturm2013spectrally}. In the RAMs, the five brightest elliptical patches correspond to the five targets. 
However, correctly locating the targets may not be an easy job based on all RAMs. In sensing, we generally calculate a threshold from a local area in the RAM
and use the threshold to tell whether a target exists or not at the range-angle grid under test. Thus, the cleaner the backgrounds of targets, the more accurately can we locate the true targets.

From Fig. \ref{fig: range angle map QPSK}(a), we see that the original communication waveform can lead to quite a strong background. Since noise is not added, the background is mainly caused by the inter-target and inter-antenna interference to be suppressed by Goals \ref{gl: low side lobes auto} and \ref{gl: small ccc}. 
Fig. \ref{fig: range angle map QPSK}(b) shows that the optimized waveform can substantially weaken the interference background over the whole range-angle domain.
This, on the one hand, highlights the necessity of optimizing the communication waveforms for better sensing; on the other hand, it shows that the optimization based on Goals \ref{gl: low side lobes auto} and \ref{gl: small ccc}
can substantially improve RAM quality. This further yields better sensing performance, as will be demonstrated in Section \ref{sec: simulations}. 

Furthermore, Fig. \ref{fig: range angle map QPSK}(c) shows that the orthogonal waveform \cite{DFRC_interleaveMIMO_OFDM_sturm2013spectrally} achieves the lowest interference background. This is expected, as the orthogonal waveform 
can fully suppress the inter-antenna interference. However, it should be noted that 
the sum rate of the orthogonal waveform \cite{DFRC_interleaveMIMO_OFDM_sturm2013spectrally} is reduced by $ 73.8 $\%\footnote{For the original and proposed waveforms, each antenna transmits $ (N-N_{\mathrm{un}}) $ data symbols, where $ N_{\mathrm{un}} $, as will be detailed in Section \ref{sec: constraint projection}, denotes the number of unused sub-carriers per antenna. 	
	The orthogonal waveform only transmits $ N/M $ data symbols per antenna. Thus, the sum rate loss is given by 	$ 1-\frac{N/M}{(N-N_{\mathrm{un}})} $. In Fig. \ref{fig: range angle map QPSK}, $ M=4 $, $ N=128 $ and $ N_{\mathrm{un}}=6 $.
}, compared with the original and proposed waveforms in Figs. \ref{fig: range angle map QPSK}(a) and \ref{fig: range angle map QPSK}(b). 
The proposed waveform optimization will incur slight loss of BER performance, as will be illustrated in Section \ref{sec: simulations}.

\subsection{Problem Formulation} \label{subsec: problem formulation}
Now that we have confirmed the feasibility of Goals \ref{gl: low side lobes auto} and \ref{gl: small ccc} in improving sensing performance, we proceed to formulate an optimization problem to realize the two goals. In doing so, we propose to moderately modify $ \mathbf{x}_m~(\forall m) $ to a level that the communication performance is only slightly affected.

Let $ \mathbf{x} $ collect all frequency-domain waveforms from the $ M $ transmitter antennas, i.e., 
\begin{align} \label{eq: x bf}
	\mathbf{x}=[\mathbf{x}_0;\mathbf{x}_1;\cdots;\mathbf{x}_{M-1}]\in \mathbb{C}^{MN\times 1},
\end{align}
where $ \mathbb{C}^{MN\times 1} $ denotes the set of $ MN\times 1 $ complex vectors, and `$ ; $' concatenates column vectors (as in MATLAB \cite{MATLAB_help}). 
{Corresponding to $ \mathbf{x} $, we let $ \mathbf{x}_{\mathrm{r}} $ denote the vector of the original communication data symbols, each drawn independently from a communication constellation, e.g., PSK or QAM.} 
Define $ \mathbf{S}_m $ as a selection matrix with the expression
\begin{align} \label{eq: Sm selection matrix}
	\mathbf{S}_m=[\mathbf{I}_M]_m^{\mathrm{T}}\otimes \mathbf{I}_N,
\end{align}
where $ [\cdot]_m $ takes the $ m $-th column of the enclosed matrix, $ \otimes $ denotes the Kronecker product, and $ \mathbf{I}_x $ is an $ x $-dimensional identity matrix.  
With the aid of $ \mathbf{S}_m $, the time-domain waveform of antenna $ m $ can be written as $ \tilde{\mathbf{x}}_m = \mathbf{F}^{\mathrm{H}}\mathbf{S}_m\mathbf{x}~(\forall m) $, where the DFT matrix $ \mathbf{F} $ is given in (\ref{eq: F bf DFT matrix}). 
Define the following circulant matrix
\begin{align}
	[\mathbf{U}_i]_{n,n'} = \left\{ 
	\substack{1,~\mathrm{if}~\myModulo{n'-n}{N}=i
		\\
		0,~\mathrm{otherwise}~~~~~
	}
	\right.,
\end{align}
where $ \myModulo{\cdot}{x} $ denotes modulo-$ x $. Then we can express the \pcc~between $ \tilde{\mathbf{x}}_m $ and $ \tilde{\mathbf{x}}_k $ as
\begin{align}\label{eq: r_mk(i)}
	& r_{mki} =  (\mathbf{F}^{\mathrm{H}}\mathbf{S}_k  \mathbf{x})^{\mathrm{H}} \mathbf{U}_i ( \mathbf{F}^{\mathrm{H}} \mathbf{S}_m  \mathbf{x} ), 
\end{align}
where $ (\cdot)^{\mathrm{H}} $ performs conjugate transposes. Note that $ m$  and $k~(m,k=0,1,\cdots,M-1) $ are indexes of transmitting antennas, and $ i~(=0,1,\cdots,N-1) $ is the index of 
range gate (or bin). Recall that the echo model (\ref{eq: y bf echo freq-domain}) is obtained based on the assumption that the target delay is no greater than the CP duration. Thus, we only focus the first $ N_{\mathrm{CP}} $ range bins, where $ N_{\mathrm{CP}} $
is the CP length.

In the case of $ m=k $, $ r_{mki} $ becomes the cyclic auto-correlation of the time-domain signal transmitted by antenna $ m $. In the case of $ m\ne k $, $ r_{mki} $ becomes the cyclic cross-correlation between signals transmitted by different antennas. 
With $ r_{mki} $ introduced, Goals \ref{gl: low side lobes auto} and \ref{gl: small ccc} can be approximated by minimizing $ \max_{\forall m,k,i\in[1,N_{\mathrm{CP}}-1]} |r_{mki}| $. The maximum can be further approximated by the $ p $-norm
with a large $ p $ \cite{MM_JXSong_weightedPeak2016TSP}. 
Therefore, the pursued waveform design is modeled as  
\begin{subequations} \label{eq: MM problem original}
	\begin{align}
		&\min_{\mathbf{x}}~ f(\mathbf{x}) = \sum_{m,k=0}^{M-1}\sum_{i=0}^{N-1} w_i |r_{mki}|^p, \label{eq: MM problem original objective}\\
		&\mathrm{s.t.}~w_i = \left\{
		\begin{array}{ll}
			1,&~\mathrm{if}~i=1,\cdots,N_{\mathrm{CP}}-1\\
			0,&~\mathrm{otherwise}
		\end{array}
		\right.;
		\label{eq: MM problem original w_i}\\
		& ~~~~~ \|\mathbf{x}\|^2=1; \label{eq: MM problem original c}\\
		&~~~~~ \mathbf{x} \approx \mathbf{x}_{\mathrm{r}}. 
		\label{eq: MM problem original d}
	\end{align}
\end{subequations}
Note that the power constraint (\ref{eq: MM problem original c})
takes unit one on the right-hand side (RHS) for convenience. When a different power value, say $ \mP $, is set, the solution to (\ref{eq: MM problem original}) only needs to be scaled by $ \sqrt{\mP} $. 
Also note that the similarity constraint (\ref{eq: MM problem original d}) is given in a general form for the moment. 
Based on the modulations used by the underlying communication system, we will re-write (\ref{eq: MM problem original d}) in more specific forms in Section \ref{sec: constraint projection}. 
As will be seen, the similarity constraint (\ref{eq: MM problem original d}) can constrain the amplitudes and phases of $ \mathbf{x} $ separately, complicating the feasible region substantially and differentiating it from a hyper-sphere \cite{Opt_hager2001minimizing_sphere}.

Problems with similar form to (\ref{eq: MM problem original}) have been established for MIMO radar waveform design \cite{MM_JXSong_weightedPeak2016TSP,MM_sequenceSetDesignJXSong2016TSP,MM_unified_LCZhao_2017TSP,book_cui2020radarWaveformOptimization}, and the MM algorithm\footnote{Briefly speaking, MM iteratively majorizes the objective function at the solution obtained in the last iteration and then minimizes the majorized function. A comprehensive tutorial of MM can be found in \cite{MM_SP_Com_ML2017TSP_Babu}.} has shown great potential in 
efficiently solving such problems like (\ref{eq: MM problem original}). 
{Enlightened by the prior success \cite{MM_JXSong_weightedPeak2016TSP,MM_sequenceSetDesignJXSong2016TSP,MM_unified_LCZhao_2017TSP,book_cui2020radarWaveformOptimization}, we choose to use the well-established MM optimization framework \cite{MM_SP_Com_ML2017TSP_Babu} for solving Problem (\ref{eq: MM problem original}). We do not claim that MM is the best option, 
but we will show that employing MM does enable us to develop an efficient algorithm for solving (\ref{eq: MM problem original}). Moreover, as widely performed in previous works \cite{MM_JXSong_weightedPeak2016TSP,MM_unified_LCZhao_2017TSP,book_cui2020radarWaveformOptimization,MM_sequenceSetDesignJXSong2016TSP}, we will first solve the problem (\ref{eq: MM problem original}) without the similarity constraint (\ref{eq: MM problem original d}) and then project the solution onto the feasible region defined by the constraint.
}

We emphasize that, despite the similarity in form, our optimization problem (\ref{eq: MM problem original}) is very different from those for MIMO radar waveform design \cite{MM_JXSong_weightedPeak2016TSP,MM_unified_LCZhao_2017TSP,book_cui2020radarWaveformOptimization,MM_sequenceSetDesignJXSong2016TSP}. 
\textit{A substantial difference lies in the expression of cyclic auto- and cross-correlations, i.e., $ r_{mki} $ given in (\ref{eq: r_mk(i)}).}
Since we optimize the frequency-domain waveforms, $ r_{mki} $ involves DFT and selection matrices; see (\ref{eq: r_mk(i)}). In contrast, the variable equivalent to $ r_{mki} $ in \cite{MM_JXSong_weightedPeak2016TSP,MM_unified_LCZhao_2017TSP,book_cui2020radarWaveformOptimization,MM_sequenceSetDesignJXSong2016TSP} do not have these matrices, as $ \mathbf{x} $ directly denotes the time-domain waveforms therein. 
This difference prevents us from directly using the existing algorithms \cite{MM_JXSong_weightedPeak2016TSP,MM_unified_LCZhao_2017TSP,book_cui2020radarWaveformOptimization,MM_sequenceSetDesignJXSong2016TSP}. 
{Furthermore, we point out that, \textit{even though the MM framework is well developed, applying it to efficiently solve
	a problem generally involves problem-specific features that require non-trivial effort to discover and exploit.} To further elaborate this point, let us consider applying the following lemma.

\mySpaceTwoMM

\begin{lemma} \label{lm: majorizing quadratic form}
	{\it The quadratic form
$ \mathbf{x}^{\mathrm{H}} \mathbf{A} \mathbf{x} $ can be majorized at $ \mathbf{x}_0 $ by $ \mathbf{x}^{\mathrm{H}} \mathbf{B} \mathbf{x} +2\Re\{ \mathbf{x}^{\mathrm{H}} (\mathbf{A}-\mathbf{B})\mathbf{x}_0 \} + \mC $, where $ \mathbf{A} $ and $ \mathbf{B}(\succeq\mathbf{A}) $ are Hermitian matrices having matching sizes with $ \mathbf{x} $ for the above calculation, and $ \mC $ absorbs irrelevant constants \cite[Lemma 1]{MM_sequenceSetDesignJXSong2016TSP}. 
}
\end{lemma}

\mySpaceTwoMM

The lemma provides a simple and useful tool for majorizing a quadratic function as will be often encountered later in Section \ref{sec: majorization}. However, applying the lemma requires us to find a suitable matrix $ \mathbf{B}(\succeq \mathbf{A}) $. 
One popular option is taking $ \mathbf{B}=\myLambdaMax{\mathbf{A}}\mathbf{I} $, where $ \myLambdaMax{\mathbf{A}} $ denotes the maximum eigenvalue of $ \mathbf{A} $. 
However, in our problem, the equivalent matrices playing the role of $ \mathbf{A} $ have the size of $ MN\times MN $, where $ M $ is the antenna number and $ N $ is the sub-carrier number. In typical IoT communication configurations, $ MN $ can be about $ 500 $ \cite{book_ahmadi2019_5G}, and hence getting $ \myLambdaMax{\mathbf{A}} $ has a computational complexity of $ \myBigO{(MN)^3}=\myBigO{10^8} $, too luxury for IoT devices. This strongly validates our point: MM indeed gives us useful tools (rules), such as Lemma \ref{lm: majorizing quadratic form}, but simply applying them does not guarantee a practically usable solution (at least in our case). 

In what follows, instead of a naive use of MM, we will disclose in Section \ref{sec: majorization} some unique signal structures and features embedded in our problem, and exploit them to majorize the objective function (\ref{eq: MM problem original objective}).
{As such, we manage to reduce $ \myBigO{10^8} $ in the above example to only $ \myBigO{10^2} $.} 
Furthermore, we will design effective projectors in Section \ref{sec: constraint projection} to deal with the similarity constraint (\ref{eq: MM problem original d}). The overall algorithm for solving (\ref{eq: MM problem original}) will be established in Section \ref{sec: over all MM algorithm}.

\section{Majorizing Objective Function (\ref{eq: MM problem original objective})} \label{sec: majorization}

According to \cite[Lemma 10]{MM_sequenceSetDesignJXSong2016TSP}, a general $ p $-norm function $ x^p $ with $ p\ge 2 $ and $ x\in[0,\bar{x}] $
can be majorized at $ x_0 $ by $ ax^2 + (px_0^{p-1}-2a x_0)x +\mC $, where 
\[
a=\myBracketRnd{\bar{x}^p-x_0^p-px_0^{p-1}(\bar{x}-x_0)}/(\bar{x}-x_0)^2.
\]
{Without obfuscation, we use $ \mC $ to absorb optimization-independent terms hereafter.}
Applying the majorization method, each summand $ |r_{mki}|^p $ given in (\ref{eq: MM problem original objective}) can be majorized individually, leading to  
\begin{align} \label{eq: f tilde (x|x(l))}
	& f(\mathbf{x})\le \tilde{f}(\mathbf{x}|\mathbf{x}^{(l)}) =\tilde{f}_1(\mathbf{x}|\mathbf{x}^{(l)}) + \tilde{f}_2(\mathbf{x}|\mathbf{x}^{(l)})+ \mC, \nonumber\\ 
	\mathrm{s.t.}~	& \tilde{f}_1(\mathbf{x}|\mathbf{x}^{(l)}) = {\sum_{m,k=0}^{M-1}\sum_{i=0}^{N-1} w_ia_{mki}^{(l)}\left| r_{mki} \right|^2};\nonumber\\
	& \tilde{f}_2(\mathbf{x}|\mathbf{x}^{(l)}) = \sum_{m,k=0}^{M-1}\sum_{i=0}^{N-1} w_i b_{mki}^{(l)}|r_{mki}|
	,
\end{align}
where $ \mathbf{x}^{(l)} $ denotes the optimal solution after the $ l $-th iteration, $ r_{mki}^{(l)} $ is obtained by substituting $ \mathbf{x}= \mathbf{x}^{(l)} $ into (\ref{eq: r_mk(i)}), 
$ |r_{mki}|^p~(\forall m,k,i) $ is majorized at $ r_{mki}^{(l)} $ over $ [0,\bar{r}^{(l)}_{mk}] $ with $ \bar{r}^{(l)}_{mk}=\max_{m,k,i}\left|r_{mki}^{(l)}\right| $,
and the two coefficients are 
\begin{align} \label{eq: a mk l i}
	a_{mki}^{(l)} = \substack{\frac{ 
			(\bar{r}^{(l)}_{mk})^p -\left| r_{mki}^{(l)} \right|^p - p\left| r_{mki}^{(l)} \right|^{p-1}
			\left(\bar{r}^{(l)}_{mk} - \left|r_{mki}^{(l)}\right| \right)
	}{\left(\bar{r}^{(l)}_{mk} - \left|r_{mki}^{(l)}\right| \right)^2}} ;
\end{align}
\begin{align} \label{eq: b mk l i}
	b_{mki}^{(l)} = p\left| r_{mki}^{(l)} \right|^{p-1} - 2a_k \left|r_{mki}^{(l)}\right|.
\end{align}
Both $ \tilde{f}_1(\mathbf{x}|\mathbf{x}^{(l)}) $ and $ \tilde{f}_2(\mathbf{x}|\mathbf{x}^{(l)}) $ in (\ref{eq: f tilde (x|x(l))}) can be further majorized.
To proceed, the following properties \cite{book_matrix2007XiandaZhang}
are useful,
\begin{gather} 
	\myVec{\mathbf{ABC}} = \myBracketRnd{\mathbf{C}^{\mathrm{T}}\otimes \mathbf{A}}\myVec{\mathbf{B}}; \label{eq: vec{ABC}=...} \\
	\myBracketRnd{\mathbf{A}^{\mathrm{H}}\otimes \mathbf{B}^{\mathrm{H}}}  \myBracketRnd{ \mathbf{C}\otimes \mathbf{D} } = \myBracketRnd{\mathbf{A}^{\mathrm{H}}\mathbf{C} } \otimes \myBracketRnd{\mathbf{B}^{\mathrm{H}} \mathbf{D} }.\label{eq: AB CD = AC BD}
\end{gather}
where $ \otimes  $ is Kronecker product, and $ \mathbf{A}, \mathbf{B} $, $ \mathbf{C} $ and $ \mathbf{D} $ are general matrices with dimensions matching for product.  

\subsection{Majorizing $ \tilde{f}_1(\mathbf{x}|\mathbf{x}^{(l)}) $ Given in (\ref{eq: f tilde (x|x(l))})} \label{subsec: majorizing f1}
We start with the following lemma to simplify the expression of $ r_{mki} $; see Appendix \ref{app: derive FUF=diag} for the proof.

\mySpaceTwoMM

\begin{lemma}\label{lm: rmk i in quadratric form}
{\it The \pcc~given in (\ref{eq: r_mk(i)}) can be simplified into
	\begin{align}%
		r_{mki} =  \mathbf{x}^{\mathrm{H}} \mathbf{A}_{mki}  \mathbf{x} , ~
		\mathrm{s.t.}~\mathbf{A}_{mki}=\mathbf{S}_k ^{\mathrm{H}} \myDiag{N[\mathbf{F}]_i^*} \mathbf{S}_m, \nonumber
	\end{align}
where $ [\mathbf{F}]_i $ denotes the $ i $-th column of the DFT matrix given in (\ref{eq: F bf DFT matrix}) and $ (\cdot)^* $ takes conjugate. 
}	
\end{lemma}

\mySpaceTwoMM

Applying (\ref{eq: vec{ABC}=...}) and Lemma \ref{lm: rmk i in quadratric form}, we can have 
\[\myVec{r_{mki}} = 
\myBracketRnd{\mathbf{x}^*\otimes \mathbf{x}}^{\mathrm{H}} \myVec{\mathbf{A}_{mki}}.\]
Since $ r_{mki} $ is already a scalar, plugging $ r_{mki} = \myVec{r_{mki}} $ 
into (\ref{eq: f tilde (x|x(l))}) turns $ \tilde{f}_1(\mathbf{x}|\mathbf{x}^{(l)}) $ into
\begin{align}\label{eq: f1 tilde x xl origin}
	& \tilde{f}_1(\mathbf{x}|\mathbf{x}^{(l)})
	=(\underbrace{\mathbf{x}^*\otimes \mathbf{x}}_{\tilde{\mathbf{x}}})^{\mathrm{H}}\myBracketRnd{ \sum_{m,k=0}^{M-1}
		\mathbf{P}_{mk} }
	\tilde{\mathbf{x}} \\ 
	&\mathrm{s.t.}~ \mathbf{P}_{mk} = \sum_{i=0}^{N-1} a_{mki}^{(l)}w_i  \myVec{\mathbf{A}_{mki}} \myVec{\mathbf{A}_{mki}} ^{\mathrm{H}}.\nonumber
\end{align}
Note that $ \tilde{f}_1(\mathbf{x}|\mathbf{x}^{(l)}) $ is in a quadratic form and, given real positive coefficients $ a_{mki}^{(l)}$ and $w_i $, $ \myBracketRnd{ \sum_{m,k=0}^{M-1}
	\mathbf{P}_{mk} } $ is Hermitian. 
Applying Lemma \ref{lm: majorizing quadratic form}, $ \tilde{f}_1(\mathbf{x}|\mathbf{x}^{(l)}) $ obtained in (\ref{eq: f1 tilde x xl origin}) can be further majorized by,
\begin{align} \label{eq: f1 tilde x xl <=}
	& \tilde{f}_1(\mathbf{x}|\mathbf{x}^{(l)})\le \tilde{\mathbf{x}}^{\mathrm{H}}\mathbf{M} \tilde{\mathbf{x}} + \nonumber\\
	&~~~~~~~~~~~~~~~ 2\Re\left\{ \tilde{\mathbf{x}}^{\mathrm{H}} \myBracketRnd{ \myBracketRnd{ \sum_{m,k=0}^{M-1}
			\mathbf{P}_{mk} } -\mathbf{M} } \tilde{\mathbf{x}}^{(l)} \right\} + \mC
		\nonumber\\
			& \mathrm{s.t.}~\mathbf{M}\succeq \myBracketRnd{ \sum_{m,k=0}^{M-1}
				\mathbf{P}_{mk} };~\tilde{\mathbf{x}}^{(l)} = \myBracketRnd{\mathbf{x}^{(l)}}^*\otimes \mathbf{x}^{(l)}.
\end{align}

The following $ \mathbf{M} $ is sufficient for the first constraint above, 
\begin{align}\label{eq: M=lambda I}
	\mathbf{M}=\bar{\lambda}\mathbf{I},~\mathrm{s.t.}~\bar{\lambda}\myDef \myLambdaMax{ \sum_{m,k=0}^{M-1}
		\mathbf{P}_{mk} }.
\end{align}
However, it is challenging to calculate $ \bar{\lambda} $, as the size of $ \myBracketRnd{ \sum_{m,k=0}^{M-1}
	\mathbf{P}_{mk} } $ can be very large. Taking $ N=100 $ and $ M=4 $ for example, the dimension of $ \mathbf{P}_{mk} $ is $ 400\times 400 $. 
The following theorem greatly simplifies the calculation of $ \bar{\lambda} $; see Appendix \ref{app: proof of theorem on bar lambda} for the proof.

\mySpaceTwoMM

\begin{theorem} \label{thrm: bar lambda}
	{\it
	The maximum eigenvalue of $ \sum_{m,k=0}^{M-1}
	\mathbf{P}_{mk} $, as denoted by $ \bar{\lambda} $, satisfies
	\begin{align}
		\bar{\lambda} = \max_{\substack{m,k=0,\cdots,M-1\\
	i=0,\cdots,N-1	
	}} N^3a_{mki}^{(l)}w_i,
	\end{align}
where $ a_{mki}^{(l)} $ is given in (\ref{eq: a mk l i}) and $ w_i $ is given in (\ref{eq: MM problem original w_i}). 
}
\end{theorem}

\mySpaceTwoMM

{Enabled by Theorem \ref{thrm: bar lambda}, $ \bar{\lambda} $ can be efficiently obtained, greatly facilitating the construction of $ \mathbf{M} $ as given in (\ref{eq: M=lambda I}). 
Combining (\ref{eq: f1 tilde x xl <=}) and (\ref{eq: M=lambda I}), we further obtain the following result; the proof is given in Appendix \ref{app: proof of corollary on f1<= final}. 

\mySpaceTwoMM

\begin{corollary} \label{col: f1 <= final}
	{\it The majorization of $ \tilde{f}_1(\mathbf{x}|\mathbf{x}^{(l)}) $ given in (\ref{eq: f1 tilde x xl <=}) can be further developed into
	\begin{align}\label{eq: f1 tilde x xl <= final}
		& \tilde{f}_1(\mathbf{x}|\mathbf{x}^{(l)}) \le \mathbf{x}^{\mathrm{H}} \myBracketRnd{
			\mathbf{Q}^{(l)}_1  - 2\bar{\lambda} \myBracketRnd{\mathbf{x}^{(l)} \myBracketRnd{\mathbf{x}^{(l)}}^{\mathrm{H}}}
		} \mathbf{x} + \mC \\
		& \mathrm{s.t.}~\mathbf{Q}^{(l)}_1 = \sum_{m,k=0}^{M-1}\sum_{i=0}^{N-1} a_{mki}^{(l)}w_i \myBracketRnd{\substack{\myBracketRnd{r_{mki}^{(l)}}^* \mathbf{A}_{mki}+~~~\\
				~~~~~~~r_{mki}^{(l)}\mathbf{A}_{mki}^{\mathrm{H}}
		}},\nonumber
	\end{align}  
	where $ \mathbf{Q}^{(l)}_1  $ is Hermitian, i.e., $ \myBracketRnd{\mathbf{Q}^{(l)}_1} ^{\mathrm{H}}=\mathbf{Q}^{(l)}_1  $.}
\end{corollary}}

\subsection{Majorizing $ \tilde{f}_2(\mathbf{x}|\mathbf{x}^{(l)}) $ Given in (\ref{eq: f tilde (x|x(l))})} \label{subsec: majorizing f2}

This majorization is relatively easier than the previous one. 
With reference to \cite[Eq. (22)]{MM_unified_LCZhao_2017TSP}, we can perform the following transformations,
\begin{align} \label{eq: |r||r (l)|>=}
	&\left|r_{mki}\right| \left|r_{mki}^{(l)}\right|\ge \myAbs{\underbrace{{r}_{mki}^* r_{mki}^{(l)}}_{x} } =\sqrt{\Re\{x\}^2+\Im\{x\}^2} \nonumber\\
	&\ge \Re\{ x \} = \frac{1}{2} \mathbf{x}^{\mathrm{H}}\myBracketRnd{ \myBracketRnd{r_{mki}^{(l)}}^* \mathbf{A}_{mki} + r_{mki}^{(l)}\mathbf{A}_{mki}^{\mathrm{H}} }\mathbf{x}.
\end{align}
As $ b_{mki}^{(l)}\le 0 $ for sure, $ \tilde{f}_2(\mathbf{x}|\mathbf{x}^{(l)}) $ given in (\ref{eq: f tilde (x|x(l))}) is then majorized by 
\begin{align}\label{eq: f2 tilde x xl <= final}
	& \tilde{f}_2(\mathbf{x}|\mathbf{x}^{(l)})\le \mathbf{x}^{\mathrm{H}} \mathbf{Q}^{(l)}_2 \mathbf{x}\\
	& \mathrm{s.t.}~\mathbf{Q}^{(l)}_2 = \sum_{m,k=0}^{M-1}\sum_{i=0}^{N-1} \frac{w_ib_{mki}^{(l)}}{2\left|r_{mki}^{(l)}\right|}\myBracketRnd{\substack{\myBracketRnd{r_{mki}^{(l)}}^* \mathbf{A}_{mki}+~~~\\
			~~~~~~~r_{mki}^{(l)}\mathbf{A}_{mki}^{\mathrm{H}}
	}}.\nonumber
\end{align}
Seen from (\ref{eq: |r||r (l)|>=}), $ \myBracketRnd{ \myBracketRnd{r_{mki}^{(l)}}^* \mathbf{A}_{mki} + r_{mki}^{(l)}\mathbf{A}_{mki}^{\mathrm{H}} } $ is Hermitian. Since $ \frac{w_ib_{mki}^{(l)}}{2\left|r_{mki}^{(l)}\right|} $ is always real, $ \mathbf{Q}^{(l)}_2 $ is Hermitian.

Substituting (\ref{eq: f1 tilde x xl <= final}) and (\ref{eq: f2 tilde x xl <= final}) into (\ref{eq: f tilde (x|x(l))}) leads to the following majorization, 
\begin{align}\label{eq: f tilde (x|x(l)) 1st final}
	&	\tilde{f}(\mathbf{x}|\mathbf{x}^{(l)}) \le \mathbf{x}^{\mathrm{H}} \myBracketRnd{
		\mathbf{Q}^{(l)}  - 2\bar{\lambda} \myBracketRnd{\mathbf{x}^{(l)} \myBracketRnd{\mathbf{x}^{(l)}}^{\mathrm{H}}}
	} \mathbf{x} + \mC\nonumber\\
	& \mathrm{s.t.}~\mathbf{Q}^{(l)} = \sum_{m,k=0}^{M-1}\sum_{i=0}^{N-1} w_ic_{mki}^{(l)} \myBracketRnd{\substack{\myBracketRnd{r_{mki}^{(l)}}^* \mathbf{A}_{mki}+r_{mki}^{(l)}\mathbf{A}_{mki}^{\mathrm{H}}
	}}; \nonumber\\
	&~~~~~ c_{mki}^{(l)} = a_{mki}^{(l)} + b_{mki}^{(l)}\Big/\myBracketRnd{ 2\left|r_{mki}^{(l)}\right| }.
\end{align}

\subsection{Further Majorizing $ \tilde{f}(\mathbf{x}|\mathbf{x}^{(l)}) $ Obtained in (\ref{eq: f tilde (x|x(l)) 1st final})} \label{subsec: majorizing f 2nd time}

Obviously, the core of the majorization function in (\ref{eq: f tilde (x|x(l)) 1st final}) is again quadratic and the matrix enclosed in the round brackets is Hermitian. Thus, we can apply Lemma \ref{lm: majorizing quadratic form} again to majorize the quadratic function, attaining
\begin{align}
& {\tilde{f}}(\mathbf{x}|\mathbf{x}^{(l)})\le \tilde{\tilde{f}}(\mathbf{x}|\mathbf{x}^{(l)})\myDef 2\Re\left\{ {\mathbf{x}}^{\mathrm{H}} \mathbf{y}^{(l)} \right\}+ {\mathbf{x}}^{\mathrm{H}}\mathbf{N} {\mathbf{x}} + \mC\nonumber\\
&%
\mathrm{s.t.}~\mathbf{y}^{(l)} = \myBracketRnd{ \mathbf{Q}^{(l)}  - 2\bar{\lambda} \myBracketRnd{\mathbf{x}^{(l)} \myBracketRnd{\mathbf{x}^{(l)}}^{\mathrm{H}}} - \mathbf{N} } {\mathbf{x}}^{(l)},
\nonumber\\
&~~~~~\mathbf{N}\succeq \mathbf{Q}^{(l)}  - 2\bar{\lambda} \myBracketRnd{\mathbf{x}^{(l)} \myBracketRnd{\mathbf{x}^{(l)}}^{\mathrm{H}}}.
\end{align} 
Similar to $ \mathbf{M} $ given in (\ref{eq: M=lambda I}), $ \mathbf{N} $ can take 
\begin{align}\label{eq: N bf >= Q bf -...}
	\mathbf{N} = \bar{\mu} \mathbf{I},~\mathrm{s.t.}~\bar{\mu}\myDef \myLambdaMax{\mathbf{Q}^{(l)}}.
\end{align}
We emphasize that a necessary condition for the above result is that the matrix $ 2\bar{\lambda} \myBracketRnd{\mathbf{x}^{(l)} \myBracketRnd{\mathbf{x}^{(l)}}^{\mathrm{H}}} $ has a non-negative minimum eigenvalue. Since $ 2\bar{\lambda} \myBracketRnd{\mathbf{x}^{(l)} \myBracketRnd{\mathbf{x}^{(l)}}^{\mathrm{H}}} $ is a rank-one matrix and $ \bar{\lambda} $ is positive for sure (according to Theorem \ref{thrm: bar lambda}), we have $ \myLambdaMin{2\bar{\lambda} \myBracketRnd{\mathbf{x}^{(l)} \myBracketRnd{\mathbf{x}^{(l)}}^{\mathrm{H}}}}=0 $ and the above necessary condition is satisfied. 

Again, we are facing a challenging problem of determining $ \bar{\mu} $. To solve it, 
we begin by disclosing some features of $ \mathbf{Q}^{(l)} $; see Appendix \ref{app: proof of lemma Q properties} for the proof.

\mySpaceTwoMM

\begin{lemma}\label{lm: Q bf l real symmetric and reformat Q bf (l)}
	{\it 
$ \mathbf{Q}^{(l)} $ is real symmetric and satisfies
\begin{align}\label{eq: Q bf (l) Kronecker product sum}
	\mathbf{Q}^{(l)} = \sum_{i=0}^{N-1} \mathbf{Q}_i^{(l)} \otimes \myDiag{[\mathbf{I}_M]_i},\mathrm{s.t.}~\mathbf{Q}_i^{(l)} = \left[ \mathbf{\Lambda}_{mki}^{(l)} \right]_{m,k=0}^{M-1},
\end{align}
where $ \mathbf{Q}_i^{(l)}~(\forall i) $ is also real symmetric and the product of any two summand matrices is a zero matrix.
}
\end{lemma}

\mySpaceTwoMM

As a result of Lemma \ref{lm: Q bf l real symmetric and reformat Q bf (l)}, 
\begin{align}\label{eq: lambdaMax Q(l)=max L{Q_i(l)}}
	\myLambdaMax{\mathbf{Q}^{(l)}} & = \max_{i=0,\cdots,N-1}  \myLambdaMax{\mathbf{Q}_i^{(l)}\otimes \myDiag{[\mathbf{I}_M]_i}} \nonumber\\
	&= \max_{i=0,\cdots,N-1}  \myLambdaMax{\mathbf{Q}_i^{(l)}},
\end{align}
where the last result is based on the fact that $ \myLambdaMax{\mathbf{A}\otimes \mathbf{B}} = \myLambdaMax{\mathbf{A}}\myLambdaMax{\mathbf{B}} $ \cite[Lemma 3]{MM_sequenceSetDesignJXSong2016TSP} given any two square matrices $ \mathbf{A} $ and $ \mathbf{B} $. 
Therefore, we turn the problem of finding the maximum eigenvalue of a very large $ MN\times MN $ matrix into finding the maximum eigenvalues of $ N $ numbers of much smaller $ M\times M $ matrices. The small matrices are fully decoupled, and hence their maximum eigenvalues can be calculated in parallel.  

\begin{algorithm}[!t]%
	\caption{\small Majorizing $ f(\mathbf{x}) $ given in (\ref{eq: MM problem original})}
	\vspace{-3mm}	
	\begin{center}\small
		\begin{tabular}{p{8.5cm}}
			\hline
			\vspace{0.5mm}
			\textit{Input}: $ \mathbf{x}^{(l)} $ (solution to the $ l $-the iteration), $ p $ (norm order), $ w_i $ given in (\ref{eq: MM problem original w_i});
			
			\begin{enumerate}[leftmargin=*]\renewcommand{\labelenumi}{{\arabic{enumi})}}
				
				\item Reshape $ \mathbf{x}^{(l)} $ into an $ N\times M $ matrix $ \mathbf{X}^{(l)} $;

				\item For each $ m $, $ \tilde{\mathbf{r}}^{(l)}_{mk} =[\mathbf{X}^{(l)}]_m\odot [{\mathbf{X}}^{(l)}]_k^*~(k=0,\cdots,M-1) $;
				
				\item Take the IDFT of $ \tilde{\mathbf{r}}^{(l)}_{mk} $, giving $ \mathbf{r}^{(l)}_{mk} $ and $ r_{mki}^{(l)} =\myBracketSqr{\mathbf{r}^{(l)}_{mk}}_i$; 
				
				\item Find $ \bar{r}^{(l)}_{mk}=\max_{m,k,i}\left|r_{mki}^{(l)}\right| $; 
				
				\item Calculate $ a_{mki}^{(l)} $ in (\ref{eq: a mk l i}), $ b_{mki}^{(l)} $ in (\ref{eq: b mk l i}) and $ c_{mki}^{(l)} $ in (\ref{eq: f tilde (x|x(l)) 1st final});
				
				\item Find $ \bar{\lambda} $ based on Theorem \ref{thrm: bar lambda};

				\item Calculate $ \mathbf{v}_{mk}^{(l)} $ in (\ref{eq: diga Lambda mk (l) = 2Re and v bf =DFT...});
				
				\item Construct $ \mathbf{Q}^{(l)} $ based on (\ref{eq: Q bf (l) in matrix}) and (\ref{eq: diga Lambda mk (l) = 2Re and v bf =DFT...});
				
				\item Construct $ \mathbf{Q}_i^{(l)}~(i=0,1,\cdots,N-1) $ as in (\ref{eq: Q bf i (l) from DFT}); 
				
				\item For each $ i=0,\cdots,N-1 $, calculate $ \myLambdaMax{\mathbf{Q}_i^{(l)}} $;

				\item 
				As per (\ref{eq: lambdaMax Q(l)=max L{Q_i(l)}}), 
				$ \bar{\mu} =  
				\max_i \myLambdaMax{\mathbf{Q}_i^{(l)}} $.

				\item Substituting $ \mathbf{x}^{(l)} $, $ \bar{\lambda} $, $ \bar{\mu} $ and $ \mathbf{Q}^{(l)} $ into (\ref{eq: f2 tilde x xl <= final}) results in the final majorization function $ \tilde{\tilde{f}}(\mathbf{x}|\mathbf{x}^{(l)})= \Re\left\{ {\mathbf{x}}^{\mathrm{H}} \mathbf{y}^{(l)} \right\} + \mC $.
				
				\vspace{-1mm}
			\end{enumerate}\\
		\end{tabular}
	\end{center}
	\label{alg: majorization}
\end{algorithm}	

The majorizations  developed in Sections \ref{subsec: majorizing f1}, \ref{subsec: majorizing f2} and \ref{subsec: majorizing f 2nd time} are summarized in Algorithm \ref{alg: majorization}. In Steps 1)-3), the \pcc s of the waveforms from the $ M $ antennas are calculated in an efficient manner. In particular, given two time-domain signals, say $ x[n] $ and $ y[n] $, their \pcc~can be calculated as IDFT$ \{X[k]Y^*[k]\} $, where $ X[k] $ and $ Y[k] $ denotes the DFTs of $ x[n] $ and $ y[n] $, respectively. Steps 5)-7) majorizes $ \tilde{f}_1(\mathbf{x}|\mathbf{x}^{(l)}) $, as developed in Section \ref{subsec: majorizing f1}. 
Steps 8) and 9) provide a fast construction of $ \mathbf{Q}^{(l)} $. In particular, based on (\ref{eq: Lamba mk l definition}), we obtain 
\begin{align} \label{eq: diga Lambda mk (l) = 2Re and v bf =DFT...}
	& \mathbf{\Lambda}_{mk}^{(l)} =\myDiag{2\Re\myBracketBig{\mathbf{v}_{mk}^{(l)}}} \nonumber\\
	&\mathrm{s.t.}~
	\mathbf{v}_{mk}^{(l)} = \mathbf{F} \left[w_ic_{mki}^{(l)}
	r_{mki}^{(l)}  \right]_{i=0}^{N-1},
\end{align} 
where $ [\cdot]_{i=0}^{N-1} $ generates an $ N $-dimensional column vector. 

Steps 10)-12) efficiently calculate $ \bar{\mu} $ based on the derivations in Section \ref{subsec: majorizing f 2nd time}. 
Note that $ \mathbf{Q}_i^{(l)} $ can be fast constructed based on $ \mathbf{v}_{mk}^{(l)} $ obtained in (\ref{eq: diga Lambda mk (l) = 2Re and v bf =DFT...}). In particular, jointly inspecting (\ref{eq: Q bf (l) Kronecker product sum}) and (\ref{eq: Q bf (l) in matrix}), we have
\begin{align} \label{eq: Q bf i (l) from DFT}
	\mathbf{Q}_i^{(l)} = \myBracketSqr{
		2\Re\{[\mathbf{v}_{mk}^{(l)}]_i \}	
	}_{m,k=0}^{M-1},
\end{align}
where the outer square brackets generate a matrix with row index $ m $ and column index $ k $. For the eigenvalue calculation in Step 12), we can apply efficient algorithms, such as the Lanczos iteration, since $ \mathbf{Q}_i^{(l)}~(\forall i) $ is a real symmetric as a result of Lemma \ref{lm: Q bf l real symmetric and reformat Q bf (l)}. Interested readers are referred to \cite{book_matrix2007XiandaZhang} for efficient eigendecomposition algorithms which shall not be further elaborated on here. 
In Step 13), we let $ \mC $ adsorb  $ {\mathbf{x}}^{\mathrm{H}}\mathbf{N} {\mathbf{x}} $ based on the structure of $ \mathbf{N} $ given in (\ref{eq: N bf >= Q bf -...}). We have also suppressed the constant scaling factor `2', as it does not affect the minimization of majorized function.

\section{Dealing With the Similarity Constraint (\ref{eq: MM problem original d})} \label{sec: constraint projection}

Based on the majorization function obtained from Step 13) in Algorithm \ref{alg: majorization}, we can recast the waveform optimization problem (\ref{eq: MM problem original}) into,
\begin{align}\label{eq: MM problem recast with majorization}
&~~~~~\mathbf{x}^{(l+1)}=\mathrm{argmin}_{\mathbf{x}}~ \Re\left\{ {\mathbf{x}}^{\mathrm{H}} \mathbf{y}^{(l)} \right\},\\
	&\mathrm{s.t.}~\mathbf{y}^{(l)} = \myBracketRnd{ \mathbf{Q}^{(l)}  - 2\bar{\lambda} \myBracketRnd{\mathbf{x}^{(l)} \myBracketRnd{\mathbf{x}^{(l)}}^{\mathrm{H}}} - \bar{\mu}\mathbf{I} } {\mathbf{x}}^{(l)}\nonumber\\
	&~~~~~\|\mathbf{x}\|^2=1~~\text{(\ref{eq: MM problem original c});   }~~~~~~  \mathbf{x}\approx \mathbf{x}_{\mathrm{r}}~~\text{(\ref{eq: MM problem original d})}, \nonumber
\end{align}
where $ \Re\left\{ {\mathbf{x}}^{\mathrm{H}} \mathbf{y}^{(l)} \right\} $ is the majorization function obtained in Step 13) of Algorithm \ref{alg: majorization}, and the two constraints given in (\ref{eq: MM problem original}) are cited here for convenience.
Without the similarity constraint (\ref{eq: MM problem original d}), the solution to problem (\ref{eq: MM problem recast with majorization}) is given by
\begin{align} \label{eq: x bf l+1 without constraint}
	\mathbf{x}^{(l+1)} = -\mathbf{y}^{(l)}/\|\mathbf{y}^{(l)}\|. 
\end{align}
Next, we develop efficient projectors to make $ \mathbf{x}^{(l+1)}  $ feasible under the similarity constraint (\ref{eq: MM problem original d}). 
We consider two representative modulations, i.e., PSK and QAM, that are widely used in standardized communication systems.

\subsection{PSK} \label{subsec: PSK projector}

We start with PSK that mainly uses phases of $ \mathbf{x}_{\mathrm{r}} $ to convey information bits. The amplitudes of $ \mathbf{x}_{\mathrm{r}} $ are less relevant and may bear more changes than phases. Therefore, the similarity constraint (\ref{eq: MM problem original d}) can be reformulated by treating the phases and amplitudes of $ \mathbf{x}_{\mathrm{r}} $ differently, as follows
\begin{gather}
	 0\le  |[\mathbf{x}_{\mathrm{r}}]_i| - |[\mathbf{x}^{(l+1)}]_i|  \le  \epa,~i=0,1,\cdots,MN-1;  \label{eq: |xi - xri|<=ea}\\
	\Big| \arg\{[\mathbf{x}^{(l+1)}]_i\}-\arg\{[\mathbf{x}_{\mathrm{r}}]_i\} \Big|\le \epp, \label{eq: |xi - xri|<=ep}
\end{gather}
where $ \arg\{\} $ takes the phase of a complex number; $ \epa $ and $ \epp $ denote the maximum tolerable changes on amplitude and phase, respectively. 
Next, we develop projectors to make $ [\mathbf{x}^{(l+1)}]_i~(\forall i) $ obtained in (\ref{eq: x bf l+1 without constraint}) satisfy (\ref{eq: |xi - xri|<=ea}) and (\ref{eq: |xi - xri|<=ep}) with minimal changes. We start with determining the feasible region defined by the two inequalities.

\begin{figure}[!t]
	\centering
	\includegraphics[width=80mm]{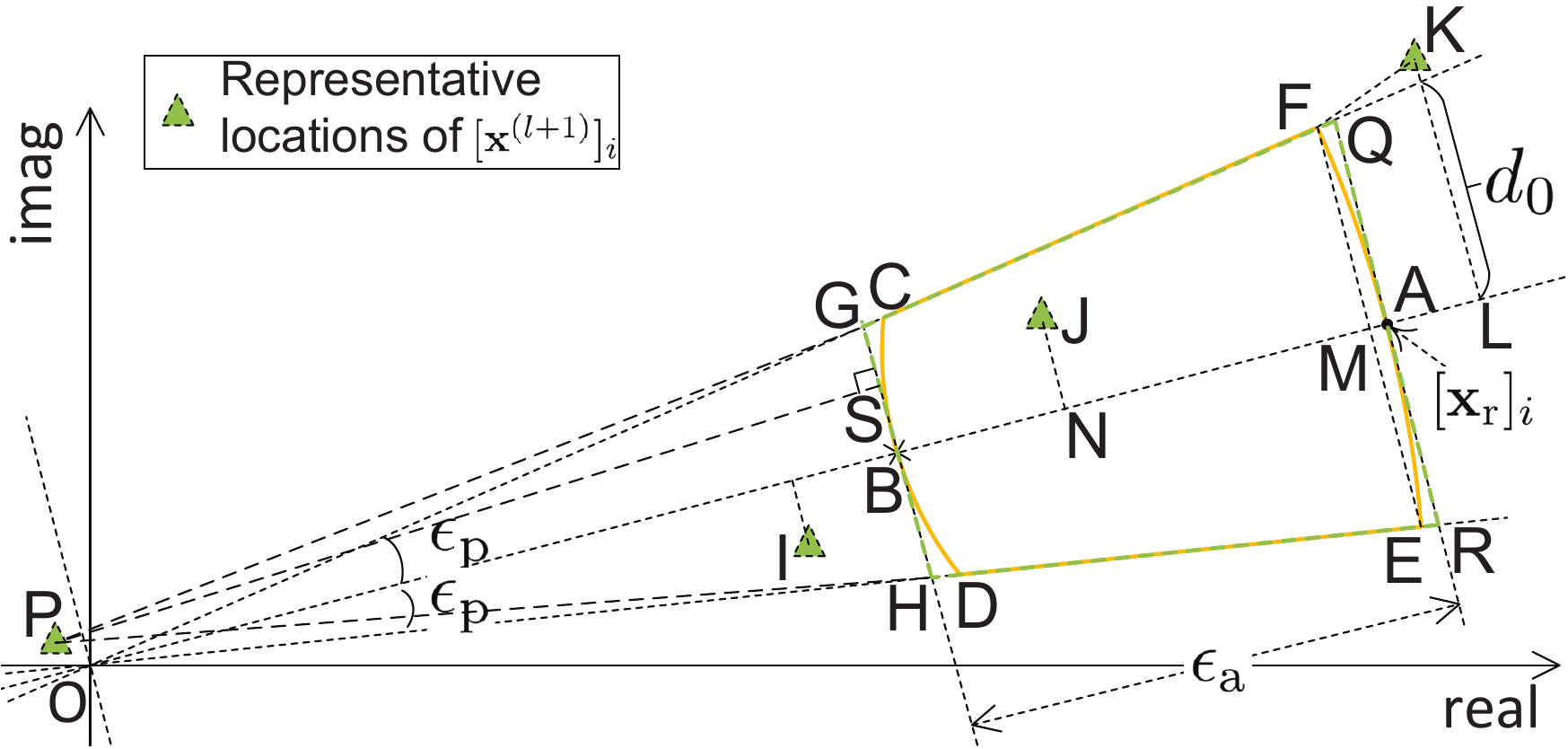}
	\caption{Geometric illustrations for developing the projector}
	\label{fig: geometric relatino for projector}
\end{figure}

A complex number can be mapped onto the two-dimensional plane spanned by the real and imaginary axes, as illustrated in Fig. \ref{fig: geometric relatino for projector}. Thus, the reference point $ [\mathbf{x}_{\mathrm{r}}]_i~(\forall i) $, as drawn from a PSK constellation, can be represented by $ \mA $ in the figure with $ |\myVecArrow{OA}|=1 $, where $ \vec{\mO\mA} $ denotes a vector from $ \mO $ to $ \mA $ and $ |\cdot| $ takes the length of a vector. The inequality in (\ref{eq: |xi - xri|<=ea}) indicates that in the direction of $ \myVecArrow{AO} $, $ |[\mathbf{x}^{(l+1)}]_i|  $ can go from $ \mA $ to $ \mB $ at most, where $ |\myVecArrow{AB}|=\epa $ as noted in Fig. \ref{fig: geometric relatino for projector}. The inequality in (\ref{eq: |xi - xri|<=ep}) implies that $ [\mathbf{x}^{(l+1)}]_i $ can only move on the circular arc $ \mE\mF $ (centered at the original) when $ |[\mathbf{x}^{(l+1)}]_i|=1 $; and it can move only on the circular arc {$ \mC\mD $} (centered at $ \mA $) when $ |[\mathbf{x}^{(l+1)}]_i-[\mathbf{x}_{\mathrm{r}}]_i|=\epa $. Based on the two extreme cases considered above, we obtain that the border of the feasible region defined by (\ref{eq: |xi - xri|<=ea}) and (\ref{eq: |xi - xri|<=ep}) is $ \mC\mD\mE\mF $. However, to simplify the projector, we replace the circular arc $ \mC\mD $ and $ \mE\mF $ with $ \mG\mH $ and $ \mQ\mR $, respectively. The two line segments are tangent to the two circular arcs. Namely, we use $ \mG\mH\mR\mQ $ as the border of the feasible region. The approximation error is negligible, as $ \epp $ is generally small.

Next, we consider three cases, as differentiated by the lengths of the projections of $ \myVecArrow{OK} $, $ \myVecArrow{OJ} $ and $ \myVecArrow{OI} $
on $ \myVecArrow{OA} $, where $ \mK $, $ \mJ $ and $ \mI $ are three representative positions of $ [\mathbf{x}^{(l+1)}]_i $ in the two-dimensional complex plane; see Fig. \ref{fig: geometric relatino for projector}. 
The three projections share the same expression, as given by $ \mathcal{P}\myVecArrow{OA} $. Here, $ \mathcal{P} $ denotes the inner product between $ \myVecArrow{OK} $ ($ \myVecArrow{OJ} $ or $ \myVecArrow{OI} $)
and $ \myVecArrow{OA} $. It can be calculated as
\begin{align} \label{eq: projection length PSK}
	\mathcal{P} = \Re\left\{  [\mathbf{x}^{(l+1)}]_i^* [\mathbf{x}_{\mathrm{r}}]_i \right\}.
\end{align} 
Note that $ \myVecArrow{OA} $ is already a unit vector; otherwise a normalization is necessary in the above expression.

\textit{Case 1}: $ \mathcal{P} >1 $. 
In this case, $ [\mathbf{x}^{(l+1)}]_i $ can be represented by $ \mK $ in Fig. \ref{fig: geometric relatino for projector}. 
Obviously,
$ 
\mK $ is outside the feasible region. How to project $ [\mathbf{x}^{(l+1)}]_i $ onto the border of the feasible region depends on the vertical distance between $ [\mathbf{x}^{(l+1)}]_i $ and the ray $ \mO\mA $. A critical vertical distance, as denoted by $ d_0 $, is achieved when $ \mK $ locates on the ray $ \mO\mF $. In Fig. \ref{fig: geometric relatino for projector}, $ |\mK\mL|> d_0 $ and we see that $ \mF $ is the closest point on the border of the feasible region to $ \mK $. Thus, the projector will replace $ [\mathbf{x}^{(l+1)}]_i $ with $ [\mathbf{x}_{\mathrm{r}}]_ie^{\mj\epp} $. Note that $ |\mK\mL|> d_0 $ can also happen when $ \mK $ locates below the ray $ \mO\mR $. Then $ \mR $, as on the border of the feasible region, will be the closest point to $ \mK $.
On the other hand, if $ |\mK\mL|\le  d_0 $, the angle of $ [\mathbf{x}^{(l+1)}]_i $ would satisfy (\ref{eq: |xi - xri|<=ep}) and we only need to scale the amplitude of $ [\mathbf{x}^{(l+1)}]_i $ to one.
Based on the geometric relation in Fig. \ref{fig: geometric relatino for projector}, we have $ d_0= \mathcal{P}\tan(\epp)
$.
Summarizing the above analyses, we obtain the following projector to make $ [\mathbf{x}^{(l+1)}]_i $ feasible in the case of $ \mathcal{P} >1 $:
\begin{align}\label{eq: projector case 1}
		& [\mathbf{x}^{(l+1)}]_i =
	\left\{
	\begin{array}{ll}
		\left.\begin{array}{ll}
			[\mathbf{x}_{\mathrm{r}}]_ie^{\mj\epp},&~\mathrm{if}~\mathcal{C}_1\\
			{}[\mathbf{x}_{\mathrm{r}}]_ie^{-\mj\epp},&~\mathrm{if}~\bar{\mathcal{C}}_1
		\end{array}\right\}
		&~\mathrm{if}~\mathcal{C}\\
		{[\mathbf{x}^{(l+1)}]_i}\Big/{|[\mathbf{x}^{(l+1)}]_i|},&~\mathrm{if}~\bar{\mathcal{C}}
	\end{array}
	\right. \\
	& \mathrm{s.t.}~\mathcal{C}:~\left| [\mathbf{x}^{(l+1)}]_i-\mathcal{P}[\mathbf{x}_{\mathrm{r}}]_i \right| >\mathcal{P}\tan(\epp)\nonumber\\
	&~~~~~ \mathcal{C}_1:~\left| [\mathbf{x}^{(l+1)}]_i - [\mathbf{x}_{\mathrm{r}}]_ie^{\mj\epp} \right| < \left| [\mathbf{x}^{(l+1)}]_i - [\mathbf{x}_{\mathrm{r}}]_ie^{-\mj\epp} \right|,\nonumber
\end{align}
where the left-hand side of the inequality in $ \mathcal{C} $ calculates the vertical distance between $ \mK $ and the ray $ \mO\mA $. Note that $ \bar{\mathcal{X}} $ takes the negation of a condition $ \mathcal{X} $. For example, $ \bar{\mathcal{C}} $ is obtained by changing `$ > $' in $ \mathcal{C} $ to `$ \le $'. 

\textit{Case 2}: $ \mathcal{P} \in [1-\epa,1] $. 
In this case, $ [\mathbf{x}^{(l+1)}]_i $ can be represented by $ \mJ $ in Fig. \ref{fig: geometric relatino for projector}. As in Case 1, the vertical distance between $ \mJ $ and the ray $ \mO\mA $ determines how to project $ [\mathbf{x}^{(l+1)}]_i $ to make it feasible. 
The critical vertical distance shares the same expression as that in Case 1
and hence is also denoted by $ d_0 $ here. 
If $ |\myVecArrow{JN}| \le d_0 $, it is feasible; otherwise, $ \mJ $ will be replaced by the intersection between $ \mJ\mN $ and $ \mG\mQ $ (or $ \mH\mR $). Based on the geometric relation shown in Fig. \ref{fig: geometric relatino for projector}, the intersection can be given by $ \frac{\mathcal{P}}{\cos(\epp)}[\mathbf{x}_{\mathrm{r}}]_ie^{\mj\epp} $ or $ \frac{\mathcal{P}}{\cos(\epp)}[\mathbf{x}_{\mathrm{r}}]_ie^{-\mj\epp} $, where $ [\mathbf{x}_{\mathrm{r}}]_ie^{\mj\epp} $ is the complex representation of $ \myVecArrow{OF} $ and $ [\mathbf{x}_{\mathrm{r}}]_ie^{-\mj\epp} $ is that of $ \myVecArrow{OE} $. To summarize, the following projector can make $ [\mathbf{x}^{(l+1)}]_i $ feasible in the case of $ \mathcal{P} \in [1-\epa,1] $:
\begin{align}\label{eq: projector case 2}
	& [\mathbf{x}^{(l+1)}]_i =
	\left\{
	\begin{array}{ll}
		\left.\begin{array}{ll}
			\frac{\mathcal{P}}{\cos(\epp)}[\mathbf{x}_{\mathrm{r}}]_ie^{\mj\epp},&~\mathrm{if}~\mathcal{C}_2\\
			\frac{\mathcal{P}}{\cos(\epp)}[\mathbf{x}_{\mathrm{r}}]_ie^{-\mj\epp},&~\mathrm{if}~\bar{\mathcal{C}}_2
		\end{array}\right\}
		&~\mathrm{if}~\mathcal{C}\\
		{}[\mathbf{x}^{(l+1)}]_i,&~\mathrm{if}~\bar{\mathcal{C}}
	\end{array}
	\right. \\
	& \mathrm{s.t.}~\mathcal{C}_2:~
	\Re\left\{  [\mathbf{x}^{(l+1)}]_i^* [\mathbf{x}_{\mathrm{r}}]_ie^{\mj\epp} \right\} > \Re\left\{  [\mathbf{x}^{(l+1)}]_i^* [\mathbf{x}_{\mathrm{r}}]_ie^{-\mj\epp} \right\},\nonumber
\end{align}
where $ \mathcal{C} $ is given in (\ref{eq: projector case 1}). Note that, geometrically, the left-hand side of $ \mathcal{C}_2 $ gives the projection length of $ \myVecArrow{OJ} $ on $ \myVecArrow{OF} $ and the right-hand side is that of $ \myVecArrow{OJ} $ on $ \myVecArrow{OE} $. We remind readers that $ \mJ $, as shown to be above $ \mG\mF $ in Fig. \ref{fig: geometric relatino for projector}, can locate below $ \mH\mR $ as well, which leads to the two cases $ \mathcal{C}_2 $ and $ \bar{\mathcal{C}}_2 $ given above.

\textit{Case 3}: $ 0\le \mathcal{P} <1 $. 
In this case, $ [\mathbf{x}^{(l+1)}]_i $ can be represented by $ \mI $ in Fig. \ref{fig: geometric relatino for projector}. 
Considering the similarity of this case to Case 1, we give the following projector without 
restating the details, 
\begin{align}\label{eq: projector case 3}
	& [\mathbf{x}^{(l+1)}]_i =
	\left\{
	\begin{array}{ll}
		\left.\begin{array}{ll}
			\frac{(1-\epa)[\mathbf{x}_{\mathrm{r}}]_ie^{\mj\epp}}{\cos(\epp)},&~\mathrm{if}~\mathcal{C}_3\\
			\frac{(1-\epa)[\mathbf{x}_{\mathrm{r}}]_ie^{-\mj\epp}}{\cos(\epp)},&~\mathrm{if}~\bar{\mathcal{C}}_3
		\end{array}\right\}
		&~\mathrm{if}~\mathcal{C}\\
		{(1-\epa)[\mathbf{x}^{(l+1)}]_i}\Big/{|[\mathbf{x}^{(l+1)}]_i|},&~\mathrm{if}~\bar{\mathcal{C}}
	\end{array}
	\right. \\
	& \mathrm{s.t.}~
	\mathcal{C}_3:~\left| [\mathbf{x}^{(l+1)}]_i - \frac{(1-\epa)[\mathbf{x}_{\mathrm{r}}]_ie^{\mj\epp}}{\cos(\epp)} \right| >\nonumber\\
	&~~~~~~~~~~~~~~~~~~~~~~~~~~~~~ \left| [\mathbf{x}^{(l+1)}]_i - \frac{(1-\epa)[\mathbf{x}_{\mathrm{r}}]_ie^{-\mj\epp}}{\cos(\epp)} \right|,\nonumber
\end{align}
where $ \mathcal{C} $ is given in (\ref{eq: projector case 1}). Geometrically, $ \frac{(1-\epa)[\mathbf{x}_{\mathrm{r}}]_ie^{\mj\epp}}{\cos(\epp)} $ and $ \frac{(1-\epa)[\mathbf{x}_{\mathrm{r}}]_ie^{-\mj\epp}}{\cos(\epp)} $ are $ \myVecArrow{OG} $ and $ \myVecArrow{OH} $, respectively.

Besides the three cases discussed above, there is a special case of $ \mathcal{P}<0 $, as represented by $ \mP $ in Fig. \ref{fig: geometric relatino for projector}. In this case, we can project $ \mP $ to $ \mG $, $ \mH $ or the vertical projection point from $ \mP $ to the line segment $ \mG\mH $, as denoted by $ \mS $, whichever one is closest to $ \mP $. Note that the vertical projection point needs to within the segment; otherwise, it is infeasible. 
This further leads to three cases. If $ \myVecArrow{GP}\cdot\myVecArrow{GH}<0 $, then we know that $ \mG $ is closest to $ \mP $; if $ \myVecArrow{HP}\cdot\myVecArrow{HG}<0 $, then $ \mH $
is closet to $ \mP $; otherwise, we need to calculate the vertical projection point. Based on Fig. \ref{fig: geometric relatino for projector}, we see that $ \myVecArrow{OS}=\myVecArrow{OG}+\myVecArrow{GS} $. Note that $ \myVecArrow{GS} $ can be calculated as $ (\myVecArrow{OP}-\myVecArrow{OG})\cdot\frac{\myVecArrow{GH}}{|\myVecArrow{GH}|} $. The complex representations of $ \myVecArrow{OP} $, $ \myVecArrow{OG} $ and $ \myVecArrow{OH} $ are $ [\mathbf{x}^{(l+1)}]_i $, $ \frac{(1-\epa)[\mathbf{x}_{\mathrm{r}}]_ie^{\mj\epp}}{\cos(\epp)} $ and $ \frac{(1-\epa)[\mathbf{x}_{\mathrm{r}}]_ie^{-\mj\epp}}{\cos(\epp)} $, respectively. Thus, $ \frac{\myVecArrow{GH}}{|\myVecArrow{GH}|}=\frac{\myVecArrow{OH}-\myVecArrow{OG}}{|\myVecArrow{OH}-\myVecArrow{OG}|}=-\mj \frac{[\mathbf{x}_{\mathrm{r}}]_i}{|[\mathbf{x}_{\mathrm{r}}]_i|} $. 
The above analyses can be summarized into the following projector
\begin{align}\label{eq: projector case <0}
	& [\mathbf{x}^{(l+1)}]_i =
	\left\{
	\begin{array}{ll}
		{(1-\epa)[\mathbf{x}_{\mathrm{r}}]_ie^{\mj\epp}}\Big/{\cos(\epp)}
		&~\mathrm{if}~\mathcal{C}_4\\
		{(1-\epa)[\mathbf{x}_{\mathrm{r}}]_ie^{-\mj\epp}}\Big/{\cos(\epp)}
		&~\mathrm{if}~\mathcal{C}_5\\
	\left.\substack{{(1-\epa)[\mathbf{x}_{\mathrm{r}}]_ie^{\mj\epp}}\Big/{\cos(\epp)}
		-~~~~~~~\\	
		\mj\frac{[\mathbf{x}_{\mathrm{r}}]_i
			\left( [\mathbf{x}^{(l+1)}]_i -
		\frac{(1-\epa)[\mathbf{x}_{\mathrm{r}}]_ie^{\mj\epp}}{\cos(\epp)}  \right)
		}{|[\mathbf{x}_{\mathrm{r}}]_i|}}\right\}
		&~\mathrm{if}~\overline{\mathcal{C}_4\cup\mathcal{C}_5}
	\end{array}
	\right. \nonumber\\
	& \mathrm{s.t.}~\mathcal{C}_4:~
-\mj\frac{[\mathbf{x}_{\mathrm{r}}]_i
	\left( [\mathbf{x}^{(l+1)}]_i -
	\frac{(1-\epa)[\mathbf{x}_{\mathrm{r}}]_ie^{\mj\epp}}{\cos(\epp)}  \right)
}{|[\mathbf{x}_{\mathrm{r}}]_i|} < 0;\nonumber\\
& ~~~~ ~\mathcal{C}_5:~\mj\frac{[\mathbf{x}_{\mathrm{r}}]_i
	\left( [\mathbf{x}^{(l+1)}]_i -
	\frac{(1-\epa)[\mathbf{x}_{\mathrm{r}}]_ie^{-\mj\epp}}{\cos(\epp)}  \right)
}{|[\mathbf{x}_{\mathrm{r}}]_i|} < 0.
\end{align}

\subsection{QAM} \label{subsec: QAM projector}

Different from PSK, QAM conveys information bits using both amplitudes and phases of $ \mathbf{x}_{\mathrm{r}} $. 
Therefore, without separating phases and amplitudes as done above, we now treat each entry of $ \mathbf{x}_{\mathrm{r}} $ as a circle center and translate the similarity constraint 
(\ref{eq: MM problem original d}) into the following feasible region, 
\begin{align} \label{eq: xi-xri < epsilon...QAM}
	\Big|[\mathbf{x}^{(l+1)}]_i-[\mathbf{x}_{\mathrm{r}}]_i\Big|^2\le \epr^2,~i=0,\cdots,MN-1.
\end{align}
Geometrically, the above inequality defines a circular area centered at $ [\mathbf{x}_{\mathrm{r}}]_i $ with the radius of $ \epr $. Therefore, the following projection can be performed, making the minimum change to each entry of $ \mathbf{x}^{(l+1)} $ to ensure its feasibility: 
\begin{align} \label{eq: projector QAM}
	& [\mathbf{x}^{(l+1)}]_i=\left\{
	\begin{array}{ll}
		[\mathbf{x}_{\mathrm{r}}]_i + \epr\frac{[\mathbf{z}]_i}{|[\mathbf{z}]_i|} & ~ \mathrm{if}~|[\mathbf{z}]_i|>\epr \\
		{}[\mathbf{x}^{(l+1)}]_i & ~ \mathrm{otherwise}
	\end{array}
	\right. \nonumber\\
	&
	\mathrm{s.t.}~[\mathbf{z}]_i = [\mathbf{x}^{(l+1)}]_i - [\mathbf{x}_{\mathrm{r}}]_i ,
\end{align}
where $ \frac{[\mathbf{z}]_i}{|[\mathbf{z}]_i|} $ is the unit direction vector and $ [\mathbf{x}_{\mathrm{r}}]_i + \epr\frac{[\mathbf{z}]_i}{|[\mathbf{z}]_i|} $ is on the border of the region defined in (\ref{eq: xi-xri < epsilon...QAM}).

\subsection{Unused Sub-carriers}
In practice, there are generally some sub-carriers not used for carrying data symbols.
They can be reserved for pilot signals or just unused due to deep fading \cite{book_ahmadi2019_5G}. 
These unused sub-carriers are actually beneficial to sensing, as they are not subject to the similarity constraint and hence provide more degrees of freedom for waveform optimization. 
However, we need to constrain the amplitudes of signals carried by unused sub-carriers to prevent severe power imbalance under the power constraint (\ref{eq: MM problem original c}). To this end, we require that the signal strength of unused sub-carriers are not greater than the maximum amplitude of the communication constellation in use. 
If PSK modulation is used, the above requirement can be enforced as follows
\begin{align} \label{eq: projector PSK unsed}
	[\mathbf{x}^{(l+1)}]_i = \frac{[\mathbf{x}^{(l+1)}]_i}{|[\mathbf{x}^{(l+1)}]_i|},~\mathrm{if}~|[\mathbf{x}^{(l+1)}]_i|>1,~\forall i\in \mathcal{I}_{\mathrm{un}},
\end{align}
where $ \mathcal{I}_{\mathrm{un}} $ collects the indexes of unused sub-carriers of the $ M $ antennas. 
If QAM is used, the maximum amplitude of either the real or imaginary part a QAM constellation point is $ \sqrt{Q}-1 $, where $ Q $ denotes the modulation order. Thus we use the following projector to enforce the above-mentioned requirement on unused sub-carriers,
\begin{align} \label{eq: projector QAM unused}
	& [\mathbf{x}^{(l+1)}]_i = \frac{\sqrt{Q}-1}{d_{\mathrm{max}}} [\mathbf{x}^{(l+1)}]_i, ~\mathrm{if}~d_{\mathrm{max}}>\sqrt{Q}-1 \nonumber \\
	& \mathrm{s.t.}~d_{\mathrm{max}} = \max\left\{ \Re\{	[\mathbf{x}^{(l+1)}]_i\} ,\Im\{	[\mathbf{x}^{(l+1)}]_i\} \right\}.
\end{align}

\section{Overall Algorithm and Analysis} \label{sec: over all MM algorithm}

\begin{algorithm}[!t]%
	\caption{\small Overall algorithm for waveform optimization}
	\vspace{-3mm}	
	\begin{center}\small
		\begin{tabular}{p{8.5cm}}
			\hline
			\vspace{0.5mm}
			\textit{Input}: $ \mathbf{x}_{\mathrm{r}} $, $ \mathbf{x}^{(0)} $, $ p $, $ w_i $, $ Q $, $ \epa $, $ \rho $, $ \mathcal{I}_{\mathrm{used}} $, $ \mathcal{I}_{\mathrm{un}}$, $ L_{\mathrm{max}} $
			
			\begin{enumerate}[leftmargin=*]\renewcommand{\labelenumi}{{\arabic{enumi})}}
				
				\item Initialize $ l=0 $ and $ \mathbf{x}^{(0)}=\mathbf{x}_{\mathrm{r}} $;

				\item Run Algorithm \ref{alg: majorization} based on $ \mathbf{x}^{(l)} $ to obtain $ \mathbf{y}^{(l)} $. Based on the intermediate results after Step 4), set $ \eta_l = \max_{\forall m,k,i} w_i r_{mki}^{(l)} $;
				
				\item If $ l\ge 1 $, then check whether $ \eta_l>\eta_{l-1} $ or not. If so, stop.
				
				\item Calculate $ \mathbf{x}^{(l+1)} $ as per (\ref{eq: x bf l+1 without constraint});

				\item If PSK is used for communications:
				
				\begin{enumerate}
					\item Set $ \epp=\frac{2\pi\rho}{Q} $; 
					
					\item For each $ i\in \mathcal{I}_{\mathrm{used}} $:
					
					\begin{enumerate}
						\item Calculate $ \mathcal{P} $ given in (\ref{eq: projection length PSK});
						\item If $ \mathcal{P}>1 $, perform the projector in (\ref{eq: projector case 1});
						
						\item If $ \mathcal{P}\in [1-\epa,1] $, perform (\ref{eq: projector case 2});
						
						\item If $ \mathcal{P}\in [0,1) $, perform (\ref{eq: projector case 3});
						
						\item If $ \mathcal{P}<0 $, perform (\ref{eq: projector case <0});
					\end{enumerate}

					\item For each $ i\in \mathcal{I}_{\mathrm{un}} $, perform (\ref{eq: projector PSK unsed});

				\end{enumerate}

				\item If QAM modulation is used for communications:
				
					\begin{enumerate}
						\item Set $ \bar{\epsilon}=2\rho $;
						
						\item Perform the projector (\ref{eq: projector QAM}) on $ [\mathbf{x}^{(l+1)}]_i $ at $ \forall i\in \mathcal{I}_{used} $;
						
						\item Perform the projector (\ref{eq: projector QAM unused}) on $ [\mathbf{x}^{(l+1)}]_i $ at $ \forall i\in \mathcal{I}_{un} $;
					\end{enumerate}

				\item If $ l=L_{\mathrm{max}}-1 $,
				  stop; otherwise, go to Step 2) with $ l=l+1 $.
				
				\vspace{-1mm}
			\end{enumerate}\\

		\end{tabular}
		\vspace{-5mm}
	\end{center}
	\label{alg: whole algorithm}
\end{algorithm}

Joining the majorization in Section \ref{sec: majorization} and the minimization in Section \ref{sec: constraint projection}, we can formulate the overall optimization algorithm for solving (\ref{eq: MM problem original}). 
The overall waveform optimization is summarized in Algorithm \ref{alg: whole algorithm}. 
In the input list, $ \mathbf{x}_{\mathrm{r}}\in \mathbb{C}^{NM\times 1} $ stacks the communication data symbols carried by all sub-carriers antenna-by-antenna;  
$ \mathbf{x}^{(0)}\in \mathbb{C}^{NM\times 1} $ is the initial sequence of the algorithm; $ p $ is the norm order in (\ref{eq: MM problem original}); $ w_i $ is a boolean weight as given in (\ref{eq: MM problem original}); $ Q $ is the modulation order of either PSK or QAM; 
$ \epa $, as given in (\ref{eq: |xi - xri|<=ea}), is the maximum tolerable changes on the amplitude of each entry of $ \mathbf{x}_{\mathrm{r}} $ when PSK modulation is used; 
$ \mathcal{I}_{\mathrm{used}} $ is the index set of sub-carriers used for communications; $ \mathcal{I}_{\mathrm{un}}=\{0,1,\cdots,MN-1\}\Big/\mathcal{I}_{\mathrm{used}} $, where $ \{\}\Big/\{\} $ returns the difference of two sets; and $ L_{\mathrm{max}} $ denotes the maximum number of iterations.

The input $ \rho $ is a ratio between the single-sided tolerable phase change, i.e., $ \epp $ in Fig. \ref{fig: geometric relatino for projector}, and the minimum angular distance between constellation points when PSK modulation is used. When QAM modulation is used, $ \rho $ is the ratio between $ \epr $ and minimum distance of constellation points. 
Based on the above definitions, we have
\begin{align} \label{eq: epa and epr}
	\epp=2\pi \rho/Q,~\epr=2\rho,~\rho<0.5,
\end{align}
where $ Q $ denotes the modulation order, and the scaling coefficient two in $ \epr $ is the minimum distance of QAM constellation points ($ \pm 1,\pm3,\cdots,\pm (\sqrt{Q}-1) $ in the real or imaginary axis).

Algorithm \ref{alg: whole algorithm} is detailed next. Step 1) sets the initial waveform as the original communication waveform. 
Step 2) majorizes the objective function and obtain the peak sidelobe as denoted by $ \eta_l $. Since the waveform design is to reduce the peak sidelobe level, Step 3) stops the algorithm if the opposite happens. Step 4) returns the waveform without the similarity constraint and then the waveform is projected onto the feasible region in Step 5)/6). Note that Algorithm \ref{alg: whole algorithm} can be accelerated using the squared iterative method (SQUAREM) \cite{MM_JXSong_weightedPeak2016TSP}. The method, in essence, performs the gradient descent using a two-point step size; please refer to \cite{MM_JXSong_weightedPeak2016TSP} for more details. 
Algorithm \ref{alg: whole algorithm squarem} extends Algorithm \ref{alg: whole algorithm} by applying SQUAREM. The computational complexity (CC) of the developed algorithms is analyzed next.

\begin{algorithm}[!t]%
	\caption{\small Waveform optimization accelerated by SQUAREM}
	\vspace{-3mm}	
	\begin{center}\small
		\begin{tabular}{p{8.5cm}}
			\hline
			\vspace{0.5mm}
			\textit{Input}: $ \mathbf{x}_{\mathrm{r}} $, $ \mathbf{x}^{(0)} $, $ p $, $ w_i $, $ Q $, $ \epa $, $ \rho $, $ \mathcal{I}_{\mathrm{used}} $, $ \mathcal{I}_{\mathrm{un}}$, $ L_{\mathrm{max}} $

			\begin{enumerate}[leftmargin=*]\renewcommand{\labelenumi}{{\arabic{enumi})}}
				
				\item Initialize $ l=0 $ and $ \mathbf{x}^{(0)}=\mathbf{x}_{\mathrm{r}} $;

				\item Run Steps 2)-6) of Algorithm \ref{alg: whole algorithm} with $ \mathbf{x}^{(l)} $, giving $ \mathbf{x}_1 $ and $ \eta_l $;
				
				\item Run Steps 2)-6) of Algorithm \ref{alg: whole algorithm} with $ \mathbf{x}^{(l)}=\mathbf{x}_1 $, giving $ \mathbf{x}_2 $;
				
				\item $ \mathbf{r}=\mathbf{x}_1-\mathbf{x}^{(l)} $; $ \mathbf{v}=\mathbf{x}_2-\mathbf{x}_1 - \mathbf{r} $; $ \alpha=-\|\mathbf{r}\|/\|\mathbf{v}\| $; 
				
				\item $ \mathbf{x} = \mathbf{x}^{(l)} - 2\alpha \mathbf{r} + \alpha^2 \mathbf{v} $;
				
				\item Make $ \mathbf{x} $ feasible via Steps 5)-6) of Algorithm \ref{alg: whole algorithm}, leading to $ \tilde{\mathbf{x}} $;
				
				\item Run Steps 1)-3) of Algorithm \ref{alg: majorization} with $ \mathbf{x}^{(l)}=\tilde{\mathbf{x}} $ and set $ \tilde{\eta} = \max_{\forall m,k,i} w_i r_{mki}^{(l)} $, where $ r_{mki}^{(l)} $ is obtained in Step 4);
				
				\item If $ \tilde{\eta} >\eta_l $, take $ \alpha=(\alpha-1)/2 $ and perform Step 5)-7) above. 
				
				\item Repeat Step 8) until the condition does not hold. 
				
				\item Set $ \mathbf{x}^{(l+1)}=\tilde{\mathbf{x}} $ and $ l=l+1 $. 
				Run Step 2)-3) of Algorithm \ref{alg: whole algorithm}.
				
				\item 
				Go to Step 2), if $ l\le L_{\mathrm{max}} $.

			\end{enumerate}\\			
		\end{tabular}
		\vspace{-5mm}
	\end{center}
	\label{alg: whole algorithm squarem}
\end{algorithm}

Algorithm \ref{alg: majorization} has a CC of $ \myBigO{M^2 N\log N} $. Specifically, Step 2) has a CC of $ \myBigO{M^2N} $. Step 3) has a CC of $ \myBigO{M^2N\log N} $, where $ \myBigO{N\log N} $ is the CC of an $ N $-dimensional (inverse) fast Fourier transform (FFT). Step 4) has a complexity of $ \myBigO{M^2 N_{\mathrm{CP}}} $, where $ MN $ is total number of $ (m,k) $ and $ N_{\mathrm{CP}} $ is the number of $ i $ values. Step 5) has a CC of $ \myBigO{M^2 N_{\mathrm{CP}}} $, as each of three equations only involves computing $ M^2 N_{\mathrm{CP}} $ scalars. Step 6) also has a CC of 
$ \myBigO{M^2 N_{\mathrm{CP}}} $, as $ w_i $ is only non-zero for $ i=0,1,\cdots,N_{\mathrm{CP}}-1 $. Step 7) has a CC of $ \myBigO{N\log N} $, which applies the fast FFT. 
Steps 8) and 9) only re-arranges existing results. Step 10) has a CC of $ \myBigO{NM^3} $, as each matrix $ \mathbf{Q}_i^{(l)} $ therein is $ M\times M $. Finally, Step 11) has a CC of $ \myBigO{N} $. From the above analysis, Steps 3) and 10) dominate the overall CC. However, in IoT applications, the number of transmitting antennas, i.e., $ M $ is generally small, while that of $ N $ is $ \myBigO{10^2} $. Thus, we conclude that 
the CC of Algorithm \ref{alg: majorization} is dominated by that of Step 3), i.e., $ \myBigO{M^2 \log N} $. 

Algorithm \ref{alg: whole algorithm} has a CC of $ \myBigO{L_{\mathrm{max}} M^2 N \log N  } $, which is an upper limit due to the maximum number of iterations. 
Enforced by Step 3), the actual number of iterations can be much smaller. The core of Algorithm \ref{alg: whole algorithm} includes Step 2)-6) that have a CC of $ \myBigO{M^2 N \log N} $. In particular,  
Step 2) runs Algorithm \ref{alg: majorization} and hence has a CC of $ \myBigO{M^2 N \log N} $. Step 4) has a CC of $ \myBigO{MN} $. Step 5) has complexity of $ \myBigO{MN} $; despite that many conditions, all projectors only involve scalar operations. Similarly, Step 6) also has complexity of $ \myBigO{MN} $. The CC of Step 2) obviously outweighs the CCs of other steps.  

When the same number of iterations are performed, Algorithm \ref{alg: whole algorithm squarem} at least doubles the CC of Algorithm \ref{alg: whole algorithm}. The least amount is achieved when the backtracking in Steps 8) and 9) does not happen. The CC of the two steps cannot be determined, as how many times we need to perform them is unpredictable in practice. However, observed from our extensive simulations, Algorithm \ref{alg: whole algorithm squarem} requires much smaller number of iterations than Algorithm \ref{alg: whole algorithm}, using the same stopping criteria, as given in Steps 10) and 11) of Algorithm \ref{alg: whole algorithm squarem} and Steps 3) and 7) of Algorithm \ref{alg: whole algorithm}. As will be shown shortly, Algorithm \ref{alg: whole algorithm squarem} mostly stops after only three iterations. Furthermore, we note that 
due to the projection in Step 6), there is no theoretical guarantee on the convergence of Algorithm \ref{alg: whole algorithm squarem}. So we stop the algorithm once the objective function starts increasing (which is enforced by Step 10), and the results in the second last iteration are returned. 
As shown next, non-trivial improvements on sensing performance can be well achieved even with only three iterations\footnote{Since the results from the last iteration are abandoned, we actually only need to run Algorithm \ref{alg: whole algorithm squarem} for two iterations. }.

\section{Simulation Results}\label{sec: simulations}

Simulation results are provided in this section to validate the proposed designs. 
The following simulation settings are mainly used: the number of sub-carriers $ N=128 $, the number of transmitting antennas $ M=4 $, the CP length $ N_{\mathrm{CP}}=N/4 $, 
the initial waveform $ \mathbf{x}^{(0)} $ is set as the original communication waveform $ \mathbf{x}_{\mathrm{r}} $,
the norm order $ p=50 $, the number of unused sub-carriers per antenna is $ \myRound{0.05N} $, the number of maximum iterations $ L_{\mathrm{max}}=10 $, and the maximum amplitude error of a unit PSK constellation point is $ \epa=0.2 $; see Fig. \ref{fig: geometric relatino for projector}. 

As mentioned in Section \ref{subsec: problem formulation}, a large norm order is used to approximate the infinite norm. From extensive observations, the impact of $ p(\ge 50) $ on the proposed waveform optimization is negligible and will not be further illustrated for brevity. Moreover, we note that $ N=128 $ and $ M=4 $ are relatively small yet practical for IoT systems, as they generally have low-profile transceivers and use small-packet communications \cite{book_ahmadi2019_5G}. 
Extension to other values of $ N $ and $ M $ is straightforward and does not provide much insight. Thus, we shall not vary these two values, and will focus on demonstrating the impact of more critical parameters: $ Q $, $ \rho $, $ \epp $, $ \epa $ and constellation types.

The first set of simulations demonstrate how the proposed algorithm performs and the changes it brings to communications and sensing. QPSK is considered first, i.e., the modulation order $ Q=4 $. The ratio between the maximum phase change and the angular interval of adjacent constellation points is $ \rho=0.15 $; see (\ref{eq: epa and epr}). Other values of $ Q $ and $ \rho $ will be illustrated shortly. Moreover, we perform $ 10^3 $ independent trials, each resetting the communication waveform $ \mathbf{x}_{\mathrm{r}} $ by randomly drawing data symbols from the QPSK constellation set. Recall that $ \mathbf{x}_{\mathrm{r}} $ is a column vector stacking data symbols on all sub-carriers antenna-by-antenna. In each trial, we run Algorithm \ref{alg: whole algorithm squarem} based on the parameters specified above.

\begin{figure}[!t]
	\centering
	\includegraphics[width=88mm]{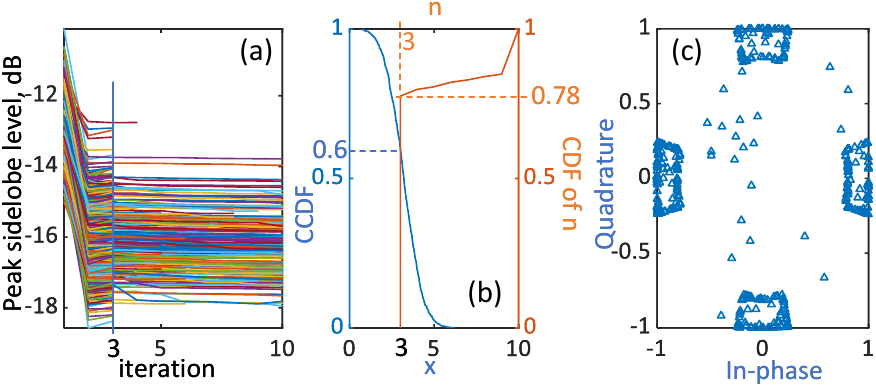}
	\caption{Subfigure (a) illustrates the value of the objective function, i.e., the peak sidelobe level, against the number of iterations, when running Algorithm \ref{alg: whole algorithm squarem} over $ 10^3 $ independent trials; (b) the left $ y $-axis shows the complementary CDF (CCDF) of the peak sidelobe improvement, as denoted by x in the figure, and the right $ y $-axis shows the CDF of the number of iterations, as denoted by n, for Algorithm \ref{alg: whole algorithm squarem} to converge; (c) demonstrates how the proposed waveform design changes the data symbols on sub-carriers of all antennas, where the symbols outside the four trapezoids are for unused sub-carriers.}
	\label{fig: mm algorithm and constellation}
\end{figure}

Fig. \ref{fig: mm algorithm and constellation}(a) plots the value of the objective function, namely the peak sidelobe level (PSL) in the auto- and cross-correlations of waveforms from communication-transmitting antennas, against the number of iterations. 
We see that the 
PSL of the original communication waveforms can be as high as about $ -10 $ dB and no less than $ -15 $ dB. 
In contrast, the proposed waveform optimization reduces 
the PSL of all $ 10^3 $ trials to below $ -12.5 $ and even reduces the lowest PSL to around $ -18.5 $ dB. 
Fig. \ref{fig: mm algorithm and constellation}(a) also shows that Algorithm \ref{alg: whole algorithm squarem} converges after three iterations for most independent trials. This is further confirmed by Fig. \ref{fig: mm algorithm and constellation}(b), where the $ y $-axis on the right is the CDF of numbers of iterations, as denoted by $ n $ in the figure. We see that $ 78 $\% independent trials have the algorithm stop after three iterations. The $ y $-axis on the left is the complementary CDF of the PSL improvement  the proposed design obtains. As highlighted in the figure, more than $ 60 $\% independent trials have their PSL improved by at least $ 3 $ dB.

Fig. \ref{fig: mm algorithm and constellation}(c) illustrates how communication waveform is changed by the proposed waveform design. Centered around the four QPSK constellation points are four trapezoid-like point clouds. They are formed by the projectors derived in Section \ref{sec: constraint projection}, which validates the effectiveness of those projectors in constraining the signal changes as desired. This result also shows the flexibility of the proposed projectors in individually constraining amplitudes and angles of constellation points. The ``untethered" points seen in Fig. \ref{fig: mm algorithm and constellation}(c) are from the unused sub-carriers. We see that their amplitudes are upper bounded well.

Earlier in Section \ref{subsec: sensing receiver processing and movitations}, Fig. \ref{fig: range angle map QPSK} in particular, we have demonstrated the improvement on the range-angle map quality brought by the proposed waveform design. Here, we further translate the improvement to extensively used performance metrics: BER for communications and the detecting probability for sensing. The communication receiving steps have been illustrated in Section \ref{subsec: cmm receiver processing}. To obtain the uncoded BER, we perform the hard decision demodulation using the signal $ \hat{\mathbf{X}} $ estimated therein. 
To evaluate detecting probability, we perform the one-dimensional cell-average constant false-alarm rate (CA-CFAR) detection along the range dimension of the obtained range-angle map. 
CA-CFAR is one of the most practically used radar detection algorithms \cite{book_richards2010principlesModernRadar}. Its implementations are briefly described as follows.

Let $ z_{mi} $ denotes the $ i $-th entry of the cyclic cross-correlation result $ \mathbf{z}_m $ obtained in (\ref{eq: z bf m}). 
CA-CFAR checks whether a target locates at the $ i $-th range grid by \textit{comparing $ |z_{mi} |^2$
with a threshold calculated based on the surrounding range grids}. The $ i $-th range grid is called the grid under test (GUT). 
The $ N_{\mathrm{gap}} $ grids adjacent to GUT are gap grids (GGs), and the $ N_{\mathrm{ref}} $ adjacent to the gap grids are the reference grids (RGs), as exemplified by
\begin{align} \label{eq: CFAR grid graph}
	\underbrace{z_{m(i-8)}\cdots z_{m(i-2)}}_{N_{\mathrm{ref}}=7\text{ RGs}},\underbrace{z_{m(i-1)}}_{\text{GG}},z_{mi},\underbrace{\cdots}_{\text{symmetric to the left}}
\end{align}
The power of the signals on the reference grids are averaged. The result is multiplied by the following $ \beta $ to produce the above-mentioned threshold \cite{book_richards2010principlesModernRadar}, 
\begin{align} \label{eq: CFAR beta coefficient}
	\beta = 2N_{\mathrm{ref}}\left( P_{\mathrm{fa}}^{-1/(2N_{\mathrm{ref}})} - 1 \right).
\end{align} 
where $ P_{\mathrm{fa}} $ denotes an expected false alarm rate. 

As shown in (\ref{eq: CFAR grid graph}), we take $ N_{\mathrm{ref}}=7 $ and $ N_{\mathrm{gap}}=1 $ for the following simulations. To calculate detecting probability, we perform $ 2\times 10^4 $ independent trials, each with randomly generated communication waveform and target ranges. 
The proposed Algorithm \ref{alg: whole algorithm squarem} is performed in each trial to generate the optimized waveform. All other parameters are kept unchanged as in the previous result, unless otherwise specified. 
Since we only perform one-dimensional CFAR, the angle information will be irrelevant. All targets then have the zero AoD to simplify echo generation.

\begin{figure}[!t]
	\centering
	\includegraphics[width=88mm]{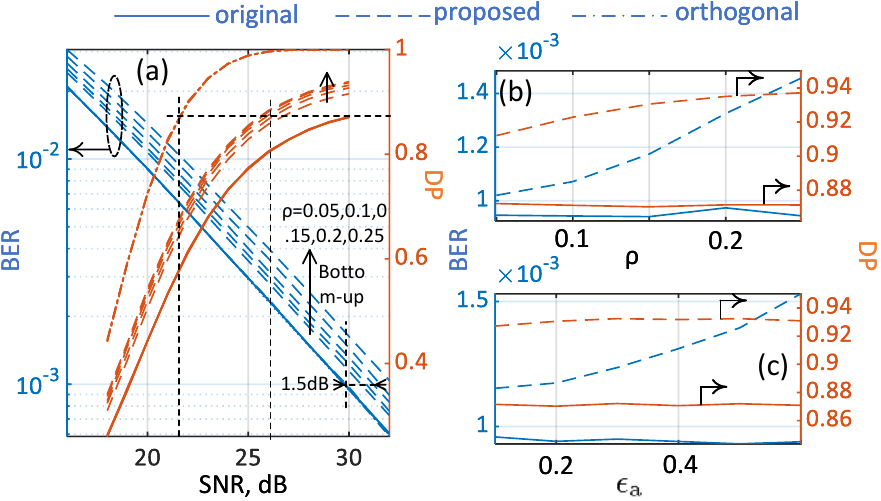}
	\caption{(a) the left $ y $-axis shows the BER performance of the original communication waveform using QPSK and the optimized one, and the right $ y $-axis shows the sensing detecting probability (DP) achieved by the original communication waveform, the optimized one and the orthogonal design \cite{DFRC_interleaveMIMO_OFDM_sturm2013spectrally}; (b) plots the BER and DP versus $ \rho $ that restricts the maximum changes on the phase of a constellation point; see (\ref{eq: epa and epr}); (c) plots the BER and DP against $ \epa $ that confines the maximum changes on the amplitude of a constellation point; see Fig. \ref{fig: geometric relatino for projector}. The legend applies to the same line styles despite colors. 
}
	\label{fig: ber pd vs snr QPSK}
\end{figure}

Fig. \ref{fig: ber pd vs snr 8_16PSK}(a) observes detecting probability against SNR, comparing the original communication waveform, the optimized one and the orthogonal one \cite{DFRC_interleaveMIMO_OFDM_sturm2013spectrally}. We take $ P_{\mathrm{fa}}=10^{-4} $ for this simulation and will show other values later. Consistent with what is observed in Fig. \ref{fig: range angle map QPSK}, the optimized waveform achieves greater detecting probabilities than the original waveform over the whole SNR region observed. More specifically, to achieve a detecting probability of $ 0.87 $, the proposed waveform design reduces the SNR by $ 4 $ dB. For the same detecting probability, the orthogonal waveform further reduces the SNR by $ 4 $ dB on top of our scheme. 

In JCAS, how much gain in sensing is generally linked to how much loss in communications.
For the orthogonal waveform, its sum rate is reduced by $ 73.8 $\%, as illustrated in Section \ref{subsec: sensing receiver processing and movitations}. 
The proposed scheme has the same sum rate as the original communications. However, our scheme can degrade the BER performance, as our sensing gain is essentially achieved by altering the data symbols on sub-carriers. 
Fig. \ref{fig: ber pd vs snr 8_16PSK}(a) shows that the BER loss increases with $ \rho $, an intermediate parameter determining the maximum angle change on a constellation point; see (\ref{eq: epa and epr}). As also shown in the figure, the loss is up to $ 2 $ dB at $ \rho=0.25 $. Note that $ \rho $ monitors the trade-off between sensing and communication performances in our design. 

Fig. \ref{fig: ber pd vs snr 8_16PSK}(b) plots detecting probability and BER against $ \rho $ to further investigates the trade-off. As expected, the detecting probability increases with $ \rho $, while the BER performances degrades increasingly with $ \rho $. However, it is interesting to notice that the increasing rate also changes with $ \rho $. In particular, the increasing rate of the detecting probability changes slowly when $ \rho $ is over $ 0.15 $.  On the other hand, the BER degrading rate for $ \rho>0.15 $ is much larger than that for $ \rho<0.15 $. 
These two results suggest that $ \rho=0.15 $ is good option, as it trades only a small BER loss for almost the maximum improvement on the detecting probability.

Fig. \ref{fig: ber pd vs snr 8_16PSK}(c) observes the detecting probability and BER performance against $ \epa $, a parameter determining the maximum amplitude change on a constellation point; see Fig. \ref{fig: geometric relatino for projector}. We see from Fig. \ref{fig: ber pd vs snr 8_16PSK}(c) that increasing $ \epa $ does not help obviously in improving sensing performance. Also, we see that the BER loss increases substantially after $ \epa $ is over $ 0.2 $. This explains why we use $ \epa=0.2 $ in Fig. \ref{fig: ber pd vs snr QPSK}(a). 
In fact, the above observation applies to other PSK modulation orders, as observed from extensive simulations. Thus, we continue using $ \epa=0.2 $ next. 

\begin{figure}[!t]
	\centering
	\includegraphics[width=88mm]{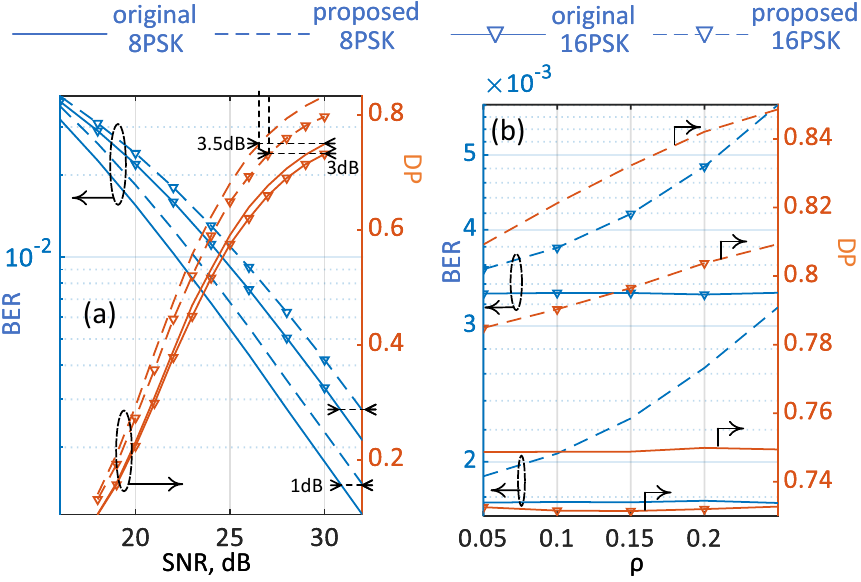}
	\caption{The BER and sensing detecting probability (DP) against SNR in (a) and versus $ \rho $ in (b). The legend applies to the same line styles despite colors.}
	\label{fig: ber pd vs snr 8_16PSK}
\end{figure}

To show the wide applicability of the proposed design, we further evaluate the BER and detecting performance of 8PSK and 16PSK with a different $ P_{\mathrm{fa}}=10^{-5} $. The results are shown in Fig. \ref{fig: ber pd vs snr 8_16PSK}. 
As stated earlier, we use $ \epa=0.2 $ and $ \rho=0.15 $. 
From Fig. \ref{fig: ber pd vs snr 8_16PSK}(a), we see that the proposed waveform optimization non-trivially improves the detecting probability over the original communication waveforms. To achieve a detecting probability of about $ 0.75 $, our design reduces the SNR requirement by $ 3.5 $ dB and $ 3 $ dB for 8PSK and 16PSK, respectively. 
The BER loss is about $ 1 $ dB for both cases. 
Fig. \ref{fig: ber pd vs snr 8_16PSK}(b) plots the BER and detecting performance against $ \rho $ for 8PSK and 16PSK. We see 
similar trends as those observed in Fig. \ref{fig: ber pd vs snr QPSK}(b) (for QPSK).

\begin{figure}[!t]
	\centering
	\includegraphics[width=88mm]{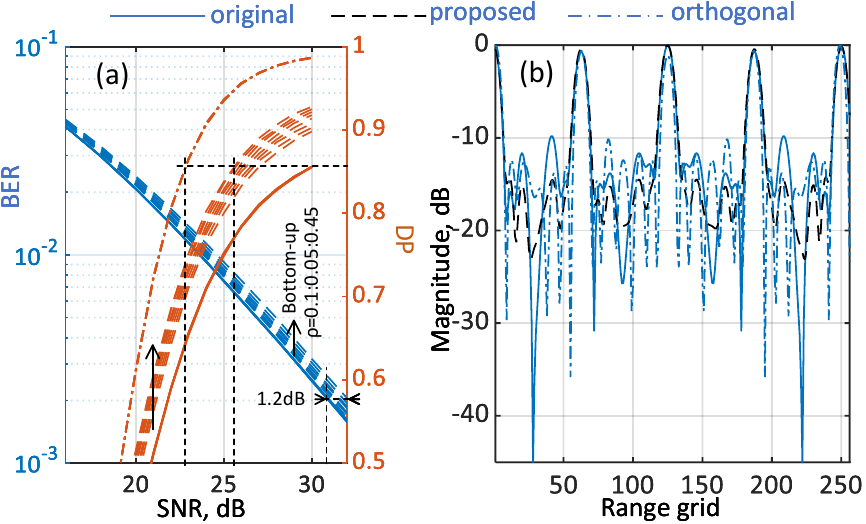}
	\caption{(a) The BER and sensing detecting probability (DP) versus SNR using the original communication waveforms based on 16QAM, the optimized one, and the orthogonal design \cite{DFRC_interleaveMIMO_OFDM_sturm2013spectrally}; (b) The range cut of the range-angle map obtained using the three types of waveforms. The legend applies to the same line styles despite colors.}
	\label{fig: ber pd vs snr qam}
\end{figure}

So far, we have shown the effectiveness of the proposed design for PSK modulations. 
Next, we further validate our design using 16QAM. Fig. \ref{fig: ber pd vs snr qam}(a) illustrates the BER and detecting performance of QAM. 
We note that $ \rho $ is different from that in Fig. \ref{fig: ber pd vs snr QPSK}. Here, it defines the radius of a circular area to confine the changes made to a QAM constellation point; see (\ref{eq: epa and epr}). We see from Fig. \ref{fig: ber pd vs snr qam}(a) that the proposed design improves the detecting performance as $ \rho $ increases. To achieve a detecting probability of about $ 0.87 $, the proposed scheme reduces the SNR requirement by up to $ 4.5 $ dB at $ \rho=0.45 $. On the other hand, the BER loss increases with $ \rho $. The maximum loss is about $ 1.2 $ dB. 

An interesting result from Fig. \ref{fig: ber pd vs snr qam}(a) is that the sensing performance gap between the orthogonal waveform and the proposed design is smaller than the gap for QPSK in Fig. \ref{fig: ber pd vs snr QPSK}. This is caused by the non-constant amplitudes of QAM constellation points. The data symbols randomly selected from the 16QAM constellation act as a random weighting when performing the matched filtering at the sensing receiver. This weighting can lead to unpredictable sidelobe changes around target peaks. In contrast, PSK constellations do not have this problem due to their constant modulus. Fig. \ref{fig: ber pd vs snr qam}(b) illustrates the range cut of the range-angle map, where five targets evenly distribute over range bins and have the zero AoD. Complied with the above illustration, the orthogonal waveform random varying sidelobes. Moreover, we see from Fig. \ref{fig: ber pd vs snr qam}(b) that the proposed scheme reduces sidelobes levels overall, hence enabling better detecting performance.

\section{Conclusions}\label{sec: conclusions}

Aiming to boost the signal availability for JCAS in IoT systems and also to achieve high-quality sensing, we propose to introduce tolerable changes to data symbols modulated on MIMO-OFDM sub-carriers to enhance the time- and spatial-domain signal orthogonality. 
Such a design is shown to reduce the inter-target and inter-antenna interference in the range-angle map. To realize the desired waveform, we establish an optimization problem and develop an efficient solution based on the MM framework. 
The high efficiency is achieved by the signal structures and features discovered by us. 
Extensive simulations are provided to validate the effectiveness of the proposed design and its superiority over the original communication waveform, specifically: 1) in general the proposed algorithm only needs a few iterations to achieve non-trivial improvement in the above-mentioned signal orthogonality; the complexity of each iteration is only $ \myBigO{M^2 N\log (N)} $ with $ M $ ($ N $) being the number of antennas (sub-carriers); 2) for 4/8/16PSK modulations, using the optimized waveform can reduce the SNR requirement by more than $ 3 $ dB to achieve the same detecting probability as using the original communication waveform, while the BER loss is less than $ 1.5 $ dB; 3) for 16QAM, the sensing SNR requirement is reduced by $ 4.5 $ dB at a cost of only $ 1.2 $ dB BER loss.

\appendix

\subsection{Deriving $ \mathbf{F} \mathbf{U}_i \mathbf{F}^{\mathrm{H}} =  \myDiag{[\mathbf{F}]_{N-i}} $}\label{app: derive FUF=diag}
The relation is based on a nice property of a circulant matrix. 
In (\ref{eq: r_mk(i)}), $ \mathbf{U}_i~(i=0,1,\cdots,N-1) $ is obviously a circulant matrix which can be diagonalized via Fourier transforms. In particular, we have  
\begin{align}\label{eq: FUF=diag}
	\mathbf{F}^{\mathrm{H}}  \mathbf{U}_i \mathbf{F} = \myDiag{N\mathbf{F}[ \mathbf{U}_i]_{1,:}^{\mathrm{T}}}.
\end{align}
Appeared in (\ref{eq: r_mk(i)}) is $ \mathbf{F}  \mathbf{U}_i \mathbf{F}^{\mathrm{H}} $
which is equal to $ (\mathbf{F}^{\mathrm{H}}  \mathbf{U}_{N-i} \mathbf{F})^{\mathrm{T}} $. 
Two relations are used for the above equality: $ \mathbf{F}^{\mathrm{T}}=\mathbf{F} $ and $ \mathbf{U}_{N-i}^{\mathrm{T}}=\mathbf{U}_i $. Based on (\ref{eq: FUF=diag}), we have 
\begin{align}
	&\mathbf{F}  \mathbf{U}_i \mathbf{F}^{\mathrm{H}} = (\mathbf{F}^{\mathrm{H}}  \mathbf{U}_{N-i} \mathbf{F})^{\mathrm{T}} = \myDiag{N\mathbf{F}[ \mathbf{U}_{N-i}]_{1,:}^{\mathrm{T}}}^{\mathrm{T}} \nonumber\\
	&=  \myDiag{[\mathbf{F}]_{N-i}},
\end{align}
where the last equality is because $ [\mathbf{U}_{N-i}]_{1,:} $ takes one at $ N-i $ and zeros elsewhere. 

\subsection{Proof of Theorem \ref{thrm: bar lambda}} \label{app: proof of theorem on bar lambda}

Based on the expression of $ \mathbf{A}_{mki} $ given in (\ref{eq: r_mk(i)}), we can apply (\ref{eq: vec{ABC}=...}) and obtain 
\begin{align} \label{eq: Pmk=SS sum SS}
	\mathbf{P}_{mk} =& \myBracketRnd{\mathbf{S}_m^{\mathrm{T}}\otimes \mathbf{S}_k^{\mathrm{H}}}\sum_{i=0}^{N-1} a_{mki}^{(l)}w_i  \myVec{\myDiag{N[\mathbf{F}]_{N-i}}}\times \nonumber\\
	&\myVec{\myDiag{N[\mathbf{F}]_{N-i}}}^{\mathrm{H}}\myBracketRnd{\mathbf{S}_m^{*}\otimes \mathbf{S}_k}.
\end{align}
Based on the expression of $ \mathbf{S}_m $ given in (\ref{eq: Sm selection matrix}), we can validate that for $ \forall (m',k')\ne (m,k) $ we have 
\begin{align} \label{eq: SS SS =0}
	\myBracketRnd{\mathbf{S}_m^{*}\otimes \mathbf{S}_k} \myBracketRnd{\mathbf{S}_{m'}^{\mathrm{T}}\otimes \mathbf{S}_{k'}^{\mathrm{H}}} = \myBracketRnd{\mathbf{S}_{m}\mathbf{S}_{m'}^{\mathrm{T}}}\otimes \myBracketRnd{\mathbf{S}_k \mathbf{S}_{k'}^{\mathrm{T}}} = \mathbf{0}.
\end{align} 
Combining (\ref{eq: Pmk=SS sum SS}) and (\ref{eq: SS SS =0}), we further have 
\begin{align}
	\mathbf{P}_{m'k'}^{\mathrm{H}}\mathbf{P}_{mk}=\mathbf{0},~\forall (m',k')\ne (m,k).
\end{align}
Since $ \mathbf{P}_{mk}~(\forall m,k) $ is Hermitian, the above result implies that the eigenvectors of $ \mathbf{P}_{mk} $ and $ \mathbf{P}_{m'k'},~\forall (m',k')\ne (m,k), $ are orthogonal. 
Then we can conclude that the eigenvectors of $ \mathbf{P}_{mk}~(\forall m,k) $ are also those of $ \myBracketRnd{ \sum_{m,k=0}^{M-1}
	\mathbf{P}_{mk} } $. 
Finally, we obtain 
\begin{align}
	\myLambdaMax{\sum_{m,k=0}^{M-1} 
		\mathbf{P}_{mk}} = \max_{\substack{m,k=0,1,\\ \cdots,M-1}} \myLambdaMax{\mathbf{P}_{mk}}. 
\end{align}
Applying the property that $ \myEigen{\mathbf{AB}}=\myEigen{\mathbf{BA}} $, $ \mathbf{P}_{mk} $ given in (\ref{eq: Pmk=SS sum SS}) satisfies
\begin{align}
	&	\myEigen{\mathbf{P}_{mk}} = \myEigen{\myBracketRnd{\mathbf{S}_m^{*}\otimes \mathbf{S}_k}\myBracketRnd{\mathbf{S}_m^{\mathrm{T}}\otimes \mathbf{S}_k^{\mathrm{H}}} \tilde{\mathbf{P}}_{mk}} \nonumber\\
	&~~~~~~~~~~~~~\myEqualOverset{(a)} \myEigen{\tilde{\mathbf{P}}_{mk}} \nonumber
	\\
	&\mathrm{s.t.}~	\tilde{\mathbf{P}}_{mk}=\sum_{i=0}^{N-1} a_{mki}^{(l)}w_i  \myVec{\myDiag{N[\mathbf{F}]_{N-i}}}\nonumber\\
	&~~~~~~~~~~~~~~~~~~~~~~~~~~~~~~~~~~	\myVec{\myDiag{N[\mathbf{F}]_{N-i}}}^{\mathrm{H}},
\end{align}
where $ \myEqualOverset{(a)} $ is because $ \myBracketRnd{\mathbf{S}_m^{*}\otimes \mathbf{S}_k}\myBracketRnd{\mathbf{S}_m^{\mathrm{T}}\otimes \mathbf{S}_k^{\mathrm{H}}}=\mathbf{I}_{MN} $.
This turns our next task into calculating the eigenvalues of $ \tilde{\mathbf{P}}_{mk} $. 

To do so, let us first investigate $ \myVec{\myDiag{N[\mathbf{F}]_{N-i}}} $. We notice that it satisfies
\begin{align}
	[\myVec{\myDiag{N[\mathbf{F}]_{N-i}}};\mathbf{0}_{N}] = N[\mathbf{F}]_{N-i}\otimes [1;\mathbf{0}_N].
\end{align}
Applying this result, we can have
\begin{align}
	& \left[\substack{\myVec{\myDiag{N[\mathbf{F}]_{N-i}}}\\ \mathbf{0}_{N}}\right]\left[\substack{\myVec{\myDiag{N[\mathbf{F}]_{N-i}}}\\ \mathbf{0}_{N}}\right]^{\mathrm{H}}= \nonumber\\
	& \left[  \substack{ \myVec{\myDiag{N[\mathbf{F}]_{N-i}}}\myVec{\myDiag{N[\mathbf{F}]_{N-i}}}^{\mathrm{H}}~~\myVec{\myDiag{N[\mathbf{F}]_{N-i}}}\mathbf{0}_N^{\mathrm{H}} 
		\\
		\mathbf{0}_N\myVec{\myDiag{N[\mathbf{F}]_{N-i}}}^{\mathrm{H}}~~\mathbf{0}_N\mathbf{0}_N^{\mathrm{H}}	
	} \right]\nonumber\\
	& = N[\mathbf{F}]_{N-i}\otimes [1;\mathbf{0}_N] \myBracketRnd{N[\mathbf{F}]_{N-i}\otimes [1;\mathbf{0}_N]}^{\mathrm{H}}\nonumber\\
	& = N^2 \myBracketRnd{[\mathbf{F}]_{N-i}[\mathbf{F}]_{N-i}^{\mathrm{H}}} \otimes \myBracketRnd{[1;\mathbf{0}_N][1;\mathbf{0}_N]^{\mathrm{H}}}.
\end{align}
This further leads to 
\begin{align} \label{eq: [P,0;0,0]}
	& \left[ \substack{\tilde{\mathbf{P}}_{mk}~~\mathbf{0}_{N^2\times N}\\
		\mathbf{0}_{N\times N^2}~~\mathbf{0}_{N\times N}
	}  \right] = N^2 \myBracketRnd{ \underbrace{\sum_{i=0}^{N-1} a_{mki}^{(l)}w_i  [\mathbf{F}]_{N-i}[\mathbf{F}]_{N-i}^{\mathrm{H}}}_{\mathbf{Q}_{mk}}
	} \nonumber\\
	& ~~~~ \otimes \myBracketRnd{[1;\mathbf{0}_N][1;\mathbf{0}_N]^{\mathrm{H}}},
\end{align}

According to \cite[Lemma 3]{MM_sequenceSetDesignJXSong2016TSP}, we have $ \myLambdaMax{\mathbf{A}\otimes \mathbf{B}} = \myLambdaMax{\mathbf{A}}\myLambdaMax{\mathbf{B}} $ given any two square matrices $ \mathbf{A} $ and $ \mathbf{B} $. Then, based on (\ref{eq: [P,0;0,0]}), we have 
\begin{align} \label{eq: last}
	\myLambdaMax{\tilde{\mathbf{P}}_{mk}} = N^2\myLambdaMax{\mathbf{Q}_{mk}}. 
\end{align}
Based on the definition of $ \mathbf{Q}_{m,k} $ given in (\ref{eq: [P,0;0,0]}), it is obvious that 
$ \mathbf{Q}_{m,k}[\mathbf{F}]_{N-i}=Na_{mki}^{(l)}w_i [\mathbf{F}]_{N-i} $. Thus, the eigenvalues of $ \mathbf{Q}_{m,k} $ is $ Na_{mki}^{(l)}w_i~(i=0,1,\cdots,N-1) $. This result, substituted into (\ref{eq: last}), leads to the expression of $ \bar{\lambda} $ given in Theorem \ref{thrm: bar lambda}. 

\subsection{Proof of Corollary \ref{col: f1 <= final}}
\label{app: proof of corollary on f1<= final}

Substituting $ \mathbf{M}=\bar{\lambda}\mathbf{I} $ into (\ref{eq: f1 tilde x xl <=}), $ \tilde{\mathbf{x}}^{\mathrm{H}}\mathbf{M} \tilde{\mathbf{x}} $ become $ \mathbf{x} $-independent under the power constraint (\ref{eq: MM problem original c}). Thus, let us focus on the first term in the second line of (\ref{eq: f1 tilde x xl <=}). In particular, based on the expressions of $ \mathbf{P}_{mk} $ and $ \tilde{\mathbf{x}} $ given in (\ref{eq: f1 tilde x xl origin}), we have %
\begin{align} \label{eq: xH P x}
	& \tilde{\mathbf{x}}^{\mathrm{H}} { 
		\mathbf{P}_{mk} } \tilde{\mathbf{x}}^{(l)}=  \sum_{i=0}^{N-1} a_{mki}^{(l)}w_i \myBracketRnd{\mathbf{x}^{\mathrm{T}}\otimes \mathbf{x}^{\mathrm{H}}} \myVec{\mathbf{A}_{mki}}\nonumber\\
	& ~~~~~~~~~~~~~~~~~\myVec{\mathbf{A}_{mki}} ^{\mathrm{H}} \myBracketRnd{\myBracketRnd{\mathbf{x}^{(l)}}^*\otimes \mathbf{x}^{(l)}} \nonumber\\
	& = \sum_{i=0}^{N-1} a_{mki}^{(l)}w_i \myBracketRnd{r_{mki}^{(l)}}^* \mathbf{x}^{\mathrm{H}}\mathbf{A}_{mki} \mathbf{x}.
\end{align}
To get the second result, the following calculations are performed:
\begin{itemize}[leftmargin=*]
	\item Applying (\ref{eq: vec{ABC}=...}) in a reverse way, we obtain 
	\begin{align} \label{eq: x otimes x vec(A)}
		\myBracketRnd{\mathbf{x}^{\mathrm{T}}\otimes \mathbf{x}^{\mathrm{H}}} \myVec{\mathbf{A}_{mki}} = \mathbf{x}^{\mathrm{H}}\mathbf{A}_{mki} \mathbf{x},
	\end{align}
	where the vectorization of the RHS is suppressed as it is already a scalar. 
	
	\item Similarly, we have 
	\begin{align} \label{eq: vec(A)^H x otimes x}
		&\myVec{\mathbf{A}_{mki}} ^{\mathrm{H}} \myBracketRnd{\myBracketRnd{\mathbf{x}^{(l)}}^*\otimes \mathbf{x}^{(l)}}  =
		\nonumber\\
		&~~~~~ \myBracketRnd{\myBracketRnd{\mathbf{x}^{(l)}}^{\mathrm{H}}\mathbf{A}_{mki} \mathbf{x}^{(l)}}^{\mathrm{H}}= \myBracketRnd{r_{mki}^{(l)}}^*,
	\end{align}
	where the second result is based on the definition of the \pcc~given in (\ref{eq: r_mk(i)}).
\end{itemize}
Similar to (\ref{eq: xH P x}), we can calculate 
\begin{align}\label{eq: xH M x}
	&	\tilde{\mathbf{x}}^{\mathrm{H}} \mathbf{M} \tilde{\mathbf{x}}^{(l)} = \bar{\lambda} \tilde{\mathbf{x}}^{\mathrm{H}}\tilde{\mathbf{x}}^{(l)} = \bar{\lambda}\myBracketRnd{
		\mathbf{x}^{\mathrm{T}}\otimes \mathbf{x}^{\mathrm{H}}
	} \myVec{ \mathbf{x}^{(l)} \myBracketRnd{\mathbf{x}^{(l)}}^{\mathrm{H}}  } \nonumber\\
	& =\bar{\lambda} \mathbf{x}^{\mathrm{H}} \myBracketRnd{\mathbf{x}^{(l)} \myBracketRnd{\mathbf{x}^{(l)}}^{\mathrm{H}}} \mathbf{x}.
\end{align}
Substituting (\ref{eq: xH P x}) and (\ref{eq: xH M x}) into (\ref{eq: f1 tilde x xl <=}) and rearranging terms, we obtain the majorization given in (\ref{eq: f1 tilde x xl <= final}). 
Note that the first term on the RHS of the inequality in (\ref{eq: f1 tilde x xl <= final}) corresponds to the real-taking term in (\ref{eq: f1 tilde x xl <=}). This implies that $ \mathbf{Q}^{(l)}_1  $ is Hermitian, i.e., $ \myBracketRnd{\mathbf{Q}^{(l)}_1} ^{\mathrm{H}}=\mathbf{Q}^{(l)}_1  $.

\subsection{Proof of Lemma \ref{lm: Q bf l real symmetric and reformat Q bf (l)}} \label{app: proof of lemma Q properties}
As the sum of two Hermitian matrices, $ \mathbf{Q}^{(l)} $ is also Hermitian.
Thus, we only need to prove the matrix is real. 
Based on the expression of $ \mathbf{A}_{mki} $ given in (\ref{eq: r_mk(i)}), $ \mathbf{Q}^{(l)} $ can be written into $ \mathbf{Q}^{(l)} = \sum_{m,k=0}^{M-1}\mathbf{S}_m^{\mathrm{H}} \mathbf{\Lambda}_{mk}^{(l)} \mathbf{S}_k $, where
\begin{align}\label{eq: Lamba mk l definition}
	\mathbf{\Lambda}_{mk}^{(l)}= \sum_{i=0}^{N-1} w_ic_{mki}^{(l)}\myBracketRnd{
		\substack{\myBracketRnd{r_{mki}^{(l)}}^* \myDiag{N[\mathbf{F}]_{N-i}} + ~\\ 
			~~~~~~ r_{mki}^{(l)} \myDiag{N[\mathbf{F}]_{N-i}}^{\mathrm{H}}}
	}. 
\end{align}
Since $ w_i$ and $c_{mki}^{(l)} $, as defined in (\ref{eq: f tilde (x|x(l)) 1st final}), are real, $ \mathbf{\Lambda}_{mk}^{(l)} $ is then a real diagonal matrix. 
Moreover, as $ \mathbf{S}_m~(\forall m) $ given in (\ref{eq: Sm selection matrix}) is real, we attain that $ \mathbf{Q}^{(l)} $ is real. 	

This can be proved by enumerating the sub-blocks of $ \mathbf{Q}^{(l)} $. In particular, with the aid of $ \mathbf{\Lambda}_{mk}^{(l)} $ and the definition of $ \mathbf{S}_m~(\forall m) $ given in (\ref{eq: Sm selection matrix}), we can express $ \mathbf{Q}^{(l)} $ as 
\begin{align} \label{eq: Q bf (l) in matrix}
	\mathbf{Q}^{(l)} = 
	\begin{pmatrix}
		\mathbf{\Lambda}_{00}^{(l)}& \mathbf{\Lambda}_{01}^{(l)}&\cdots&\mathbf{\Lambda}_{0(M-1)}(l)\\
		\mathbf{\Lambda}_{10}^{(l)}& \mathbf{\Lambda}_{11}^{(l)}& \cdots& \mathbf{\Lambda}_{1(M-1)}(l)\\
		\vdots& \vdots & \ddots & \vdots\\
		\mathbf{\Lambda}_{(M-1)0}^{(l)}&\mathbf{\Lambda}_{(M-1)1}^{(l)}&\cdots&\mathbf{\Lambda}_{(M-1)(M-1)}(l)
	\end{pmatrix} 
\end{align}
which then leads to (\ref{eq: Q bf (l) Kronecker product sum}). 	
Applying the property given in (\ref{eq: AB CD = AC BD}), we have
\begin{align}
	& \myBracketRnd{\mathbf{Q}_i^{(l)} \otimes \myDiag{[\mathbf{I}_M]_i}} \myBracketRnd{\mathbf{Q}_{j}^{(l)} \otimes \myDiag{[\mathbf{I}_M]_j}} \nonumber\\
	&= \myBracketRnd{\mathbf{Q}_i^{(l)}\mathbf{Q}_{j}^{(l)}}\otimes \myBracketRnd{\myDiag{[\mathbf{I}_M]_i}\myDiag{[\mathbf{I}_M]_j}} = \mathbf{0}. \nonumber
\end{align}
According to Lemma \ref{lm: Q bf l real symmetric and reformat Q bf (l)}, we have $ \mathbf{\Lambda}_{mki}^{(l)}=\mathbf{\Lambda}_{kmi}^{(l)} $. This makes $ \mathbf{Q}_i^{(l)}~(\forall i) $ given in (\ref{eq: Q bf (l) Kronecker product sum}) also real and symmetric.

\bibliographystyle{IEEEtran}
\bibliography{IEEEabrv,./bib_JCAS.bib}

\end{document}